\documentclass[preprint,floatfix,aps,prd,showpacs,footinbib,amsmath,amssymb,amsfonts,superscriptaddress]{revtex4-1}
\usepackage{bm}
\usepackage{graphicx}
\usepackage{color}
\usepackage{dsfont}
\usepackage[normalem]{ulem}
\usepackage{hyperref}
\usepackage{mathrsfs}
\usepackage{caption}
\DeclareCaptionLabelFormat{adja-page}{\hrulefill\\#1 #2 \emph{(previous page)}}
\usepackage{subfig}
\usepackage{calrsfs}
\usepackage[utf8]{inputenc}
\usepackage{varioref}
\usepackage[active]{srcltx}

\expandafter\ifx\csname package@font\endcsname\relax\else
\expandafter\expandafter
\expandafter\usepackage
\expandafter\expandafter
\expandafter{\csname package@font\endcsname}%
\fi
\hypersetup{
	colorlinks=true,       
	linkcolor=blue,          
	citecolor=blue,        
}

\usepackage[section]{placeins}

\usepackage{fancybox}
\labelformat{figure}{Fig.~#1} 

\begin{document}
\title{Quasinormal modes as a distinguisher between general relativity and $ f(R) $ gravity:  Charged black-holes}


\author{Soham Bhattacharyya} 
\affiliation{School of Physics, Indian Institute of Science Education and Research Thiruvananthapuram (IISER-TVM), Trivandrum 695551, India}
\email{xeonese13@iisertvm.ac.in}
\author{S. Shankaranarayanan}
\affiliation{Department of Physics, Indian Institute of Technology Bombay, Mumbai 400076, India}
\email{shanki@phy.iitb.ac.in}

\begin{abstract}
The ring-down phase of black-hole perturbations is governed by the Quasi-Normal modes (QNM) and offer valuable insight into the nature of the objects emitting them, raising an interesting question: Whether QNMs can be used to distinguish between theories of gravity? 
We construct a consistency test of General Relativity (GR) which enables one to distinguish between General relativity and a specific class of modified theories of gravity: $ f(R) $. We show that an energetic inequality between scalar (polar) and vector (axial) type gravitational perturbations will exist for Reissner-N\"{o}rdstrom solutions of GR - using which we find a novel method of determining the charge of a non-spinning black hole in GR. We then show that there will be a further energetic difference for charged black holes in $ f(R) $. Finally, we utilize this extra difference to construct a parameter to quantify deviation from GR.
\end{abstract}


\maketitle

\section{Introduction}

General Relativity has been an unprecedented success in describing the majority of astrophysical phenomena. Its recent successes include the direct detection of gravitational waves from the binary black hole and binary neutron star mergers by the LIGO-VIRGO collaboration~\cite{Abbott:2016blz,*Abbott:2017oio,TheLIGOScientific:2017qsa}. These detections have raised questions about the possibility of obtaining the severe constraints on the degree of validity of general relativity in the strong gravity regimes. 

General relativity is far from being a complete theory. It predicts black-holes, with a singularity at their centers, and represents a breakdown of general relativity as the classical description cannot be expected to remain valid in the extreme condition near the singularity. Physically, the singularity of the stationary vacuum isolated black-hole solutions is connected with the infinite growth of the curvature invariants, such as the Kretschmann invariant. 

Several modifications to general relativity have been proposed to remove the singularity. In general, the modifications contain higher derivative and nonlocality~\cite{Stelle1977,Clifton2012,Will:1993ns,DeFelice:2010aj,RevModPhys.82.451,Nojiri:2010wj,*Nojiri:2017ncd}. The question that naturally arises is, how to distinguish between general relativity and modified theories of gravity? Are there any unique signatures for the modified gravity theories that can potentially be detected in the terrestrial (like Einstein Telescope) or space based observations (like eLISA)?

Astrophysical black-holes that are in the centers of the galaxies or formed due to the collapse interact with the external surroundings. Thus, these black-holes are perturbed continuously compared to the
exact solutions in general relativity or modified theories of gravity.  The perturbed black-hole responds by emitting gravitational waves~\cite{Wheeler,Zerilli1970a,*Zerilli1970,*Zerilli1974,Chandrasekhar:1985kt,Vishveshwara1970a}. More specifically, the response consists of a broadband burst, followed by the quasi-normal mode ringing \cite{Mino:1997bx,Sasaki:2003xr,Kokkotas:1999bd,Berti:2009kk,0264-9381-16-12-201,Konoplya:2011qq}. Interestingly, the quasi-normal modes --- damped resonant modes of black-holes --- are independent of the nature of the perturbation and, only depends on the black-hole parameters like Mass $M$, Charge $Q$ and angular momentum $a$~\cite{Wheeler,Zerilli1970a,*Zerilli1970,*Zerilli1974,Chandrasekhar:1985kt,Vishveshwara1970a}. 

Many modified theories of gravity predict an extra degree of freedom, besides the two transverse modes, in the gravitational waves~\cite{Berry2011,Capozziello2008,PhysRevD.50.6058,*Abbott:2017oio,Myung2016,Capozziello:2017vdi}. The question that raises is: Whether the extra mode leaves any signatures in the quasi-normal mode ringing and provide a new way of distinguishing general relativity from modified theories of gravity? Previous analysis of Reissner-N\"{o}rdstrom black holes in $ f(R) $ theories were done in \cite{Nojiri:2014jqa,*Nojiri:2017kex,DeLaCruz-Dombriz2009}

Recently the current authors showed explicitly while the two types of black hole perturbations --- scalar (polar) and vector (axial) --- share equal amounts of emitted gravitational energy in General relativity, in $f(R)$ theories they do not share same amounts of emitted gravitational energy. The current authors also identified a parameter to distinguish between general relativity and $f(R)$ \cite{Bhattacharyya2017}. 

In this work, we extend the analysis for charged Reissner-N\"ordsrtom (RN) black-holes in $f(R)$ theories whose action is:
\begin{equation}\label{fRactionI}
S= \int d^{4}x \, \sqrt{-g} \, \left[\frac{f(R)}{2 \kappa^2}+\mathcal{L}_{EM}\right] \, ;\qquad \kappa^2=\frac{8\pi G}{c^4} \, ,
\end{equation}
where 
\begin{equation}
\mathcal{L}_{EM} = -\frac{1}{4}F^{\alpha\beta}F_{\alpha\beta} 
\end{equation}
is the Lagrangian density for classical electrodynamics. Although, $f(R)$ theories are higher derivative theories, they do not 
suffer from Ostr\"ogradsky instability \cite{Woodard2007}. They arise as a low-energy limit of many superstring theories \cite{Starobinsky1980,Zwiebach:1985uq,TSEYTLIN198692,Oh:1985bm,Charmousis:2008kc}. Also, higher-order Ricci scalar terms can account for some high-energy modifications.
 
It is important to note that the astrophysical black-holes probably neutralize their electric charge rather quickly and are expected to remain nearly neutral. However, the reasons for the doing the detailed analysis for Reissner-N\"{o}rdstrom black-hole are: (i) Like Kerr, unlike Schwarzschild, Reissner-N\"{o}rdstrom has two parameters to describe the black-hole. (ii) Unlike Kerr, Reissner-N\"{o}rdstrom is spherically symmetric, and it gives critical insight into obtaining a quantifying tool for deviations from general relativity. (iii) The time-scales involved in charge neutralization is usually longer than the time-scale in which black-hole forms in an NS-NS merger \cite{TheLIGOScientific:2017qsa}. Parameter estimation of these black holes are done by matching the observed waveforms with available simulation templates. However, due to the immense computing power already required to estimate such parameters, currently there is no template for obtaining the charge these black holes might possess. In this work, we show that QNMs at the epoch of the formation of such black-hole will contain signatures of the charge of these black-holes. More specifically, we show that it is possible to estimate the charge of a black hole from the energetics of quasi-normal modes. We also explicitly obtain a measure for deviations from general relativity using the energetics of quasinormal modes in $f(R)$ theories.

In Section \ref{sec:RNinGR}, we formulate the problem of a charged black-hole perturbations in general relativity and illustrate a novel method of finding the charge of a Reissner-N\"{o}rdstrom black-hole --- an unique method proposed in this work. In Section \ref{fR}, we obtain an expression for the radiated energy-momentum of perturbation for a charged black-hole space-time. In Section \ref{quantfR} we extend the gauge invariant analysis of Section \ref{sec:RNinGR} for a charged black hole in $ f(R) $ theory and obtain a new quantifying measure of the energetic difference due to the presence of the massive scalar mode.

In this work the metric signature we adopt is $(-,+,+,+)$ and 
we set $ G=c=1 $, $ 4\pi\epsilon=1 $, implying that a point charge $ Q $ has a Coulomb potential $\frac{Q}{r}$. We use Greek letters to refer to 4-dimensional space-time indices $(0 \cdots 3)$, lower Latin indices refer to the orbit space coordinates $(0,1)$, and upper Latin for the two angular coordinates $(2,3)$. The various physical quantities with the {\it over-line} refer to the values evaluated for the spherically symmetric background, whereas superscript $(n)$ represents the n-th order perturbed quantity.

\section{Perturbations of RN space-time in general relativity}\label{sec:RNinGR}

In this section, we discuss the formalism to obtain the linear order perturbations about Reissner-N\"{o}rdstrom space-time in general relativity (GR). The discussion in this section follows the formalism developed by Kodama and Ishibashi~\cite{Kodama2000,*Kodama2003,*Kodama2004}. In Sec. \ref{enGR}, we show that for a non-spinning black-hole in GR, any energetic difference would imply that the black-hole possesses charge.

\subsection{Dynamics of perturbation} \label{grdyn}

Following Refs. \cite{Kodama2000,*Kodama2003,*Kodama2004}, we decompose the 4-D background space-time with metric $ \overline{g}_{\mu\nu}$ into a product of an orbit space with metric $ g_{ab} $ and a 2-sphere $ (\mathcal{S}^2) $ with metric $ \gamma_{AB} $. The background line element is \cite{Kumar:2015bha}
\begin{eqnarray}
ds^2 & =&  -g(y)dt^2+\frac{1}{g(y)}dr^2+ \rho^2(y) \, d\Omega^2 \label{bgm}
\end{eqnarray}
where 
\begin{eqnarray}
& & g(y)=1-\frac{1+q^2}{y}+\frac{q^2}{y^2}; \rho(y)=y=\frac{r}{r_H} \nonumber \\
& & M=\frac{r_H(1+q^2)}{2}; Q=q \, r_H; r_H=M+\sqrt{M^2-Q^2} \, .
\end{eqnarray} 
$r_H$ is the horizon radius, $y$ is the dimensionless radius from the center, $M$ is mass of the black-hole, and $q$ is the dimensionless charge of the hole, scaled by the horizon radius $r_H$. 

Unlike Schwarzschild and Kerr, RN space-times are characterized by a non-zero background energy-momentum tensor corresponding to the electromagnetic field tensor
\begin{eqnarray}
\bar{T}_{\mu\nu} &=& \bar{F}_{\mu\alpha}\bar{F}_\nu\,^\alpha\ \label{emem} -\frac{1}{4}\bar{g}_{\mu\nu}\bar{F}_{\alpha\beta}\bar{F}^{\alpha\beta} \, .
\end{eqnarray}
However, $ \overline{g}_{\mu\nu} \overline{T}^{\mu\nu}=0 $ and hence, the space-time is Ricci flat. The background equations of motion are given by
\begin{eqnarray}
\overline{R}_{\mu\nu} &=& \kappa^2 \overline{T}_{\mu\nu} \\
\overline{F}^{\mu\nu}\,_{;\mu} &=& 0 \\
\overline{T}^{\mu}_{\nu} &=& \left(\begin{array}{c|c}
-P\delta^a_b & 0\\
---&---       \\
0 & P\delta^A_B
\end{array}\right) \\
P &=& \frac{Q^2}{\kappa^2y^4} \, .
\end{eqnarray}
Perturbations about a curved background space-time $ \bar{g}_{\mu\nu} $ can be represented as
\begin{eqnarray}
\label{eq:metricpert}
g_{\mu\nu} &=& \overline{g}_{\mu\nu} + \epsilon \, h_{\mu\nu}
\end{eqnarray}
where $ \epsilon $ is a book-keeping parameter which will be set to unity at the end of the calculation. Similarly,
\begin{eqnarray}
g^{\mu\nu} &=& \overline{g}^{\mu\nu} - \epsilon \, h^{\mu\nu} + \mathcal{O}(\epsilon^2)\\
R_{\mu\nu} &=& \overline{R}_{\mu\nu} + \epsilon \, R^{(1)}_{\mu\nu} + \epsilon^2 R^{(2)}_{\mu\nu}
\end{eqnarray}

Although the perturbations $ h_{\mu\nu} $ is a tensor in the full 4-dimensional space-time, individual components behave differently under rotations in the subspace ($ \mathcal{S}^2 $). Specifically, $ h_{ab} $ transform as scalars, $ h_{aB} $ as vectors, and $ h_{AB} $ as tensors, thus enabling us to separate $ h_{\mu\nu} $ into scalar, vector, and tensor parts, respectively. As shown in 
Ref.~\cite{Bhattacharyya2017}, tensor perturbations do not exist in $(2+2)$ space-times, hence, $ h_{\mu\nu} $ separates into scalar and vector parts which are decoupled at linear order as
\begin{eqnarray}\label{metdecomp}
h_{\mu\nu} & =& h^V_{\mu\nu}+h^S_{\mu\nu} \, .
\end{eqnarray}
All the components of the perturbation tensors $ h_{\mu\nu} $ are not independent. Two gauge invariant functions - $ \Phi^0_V $ and $ \Phi^0_S $, from the components of $ h_{\mu\nu}^{V/S} $ completely determine the gravitational sector of the perturbation \cite{Kodama2004b}.

For the electromagnetic field,  the electromagnetic four-potential corresponding to the background line-element (\ref{bgm}) is given by 
\begin{equation}
\bar{A}_\mu\equiv\left(A_a,A_A\right)=\left(\frac{\sqrt{2}Q}{\kappa r},0,0,0\right) \, .
\end{equation}
Under coordinate transformations in $ \mathcal{S}^2 $, $ A_a $ transform as scalars while $ A_A $ transform as vectors \cite{Kodama2004}.  Using this property, the perturbed Electromagnetic field tensor $ \delta F_{\mu\nu} $ can be 
similarly separated into
\begin{eqnarray}
\delta F_{\mu\nu} & =& \delta F^V_{\mu\nu}+\delta F^S_{\mu\nu},
\end{eqnarray}
at linear order. Gauge invariant master variables $ \mathcal{A}_V $ and $ \mathcal{A}_S $ defined from $ \delta F_{\mu\nu} $ and the perturbed Maxwell equations ~\cite{*Kodama2004} similarly classify the Electromagnetic sector of the perturbations.

The perturbation equations for each $\ell$, after assuming a time dependence of $ e^{i\tilde{\omega} \tilde{t}} $ ($ \tilde{\omega}=r_H\omega $ and $ \tilde{t}=\frac{t}{r_H} $), satisfy the following four effective one-dimensional second order differential equations: 
\begin{eqnarray}
\frac{d^2\Phi^V_{\pm}}{dx^2}+\left(\tilde{\omega}^2-V^V_{\pm}\right)\Phi^V_{\pm} & =& 0 \label{GRpert1}\\
\frac{d^2\Phi^S_{\pm}}{dx^2}+\left(\tilde{\omega}^2-V^S_{\pm}\right)\Phi^S_{\pm} & =& 0 \, , \label{GRpert2} 
\end{eqnarray}
where 
\begin{equation}
 x=y-\frac{\ln (y-1)}{q^2-1}+\frac{q^4 \ln \left(y-q^2\right)}{q^2-1}
\end{equation}
is the generalized, scaled \textit{tortoise coordinate}, $ V_{\pm}^{V/S} $ are short-range scattering potentials (defined for each $ \ell\geq2 $) with a maximum and asymptotically falling to zero for $ x\rightarrow\pm\infty $. Unlike the Schwarzschild \cite{Bhattacharyya2017}, the above quantities $ \Phi^V_{\pm}, \Phi^S_{\pm}$ are the superposition of the gravitational and electromagnetic gauge-invariant perturbations. Details can be seen in Appendix \ref{grkod}.

\subsection{Difference in radiated energy flux at infinity between scalar and vector perturbations}\label{enGR}
 
In the case of Schwarzschild, only gravitational (scalar and vector) modes exist. The scalar and vector modes are related. Also, the effective potentials they satisfy are associated.  Thus, gravitational radiation from the perturbed Schwarzschild black-hole, as detected at asymptotic spatial infinity, have an equal contribution from the scalar and vector modes \cite{Chandrasekhar:1985kt,Bhattacharyya2017}. 

In the case of RN, as mentioned above, along with the two gravitational modes, two types of electromagnetic perturbations are also present. As discussed in Appendix \ref{grkod}, it is the superposition of the gravitational and electromagnetic perturbations that are related. Thus, the two perturbations cannot be treated separately; perturbation of one will affect the other~\cite{*Kodama2004,Gunter1980,Chandrasekhar:1985kt}. In other words, an incident wave from radial infinity in the RN space-time that is purely gravitational will result in a scattered wave which has both gravitational and electromagnetic component. 

Gunther~\cite{Gunter1980} quantified the fraction of incident gravitational radiation converted into electromagnetic radiation due to the scattering process.  Defining the conversion factors $ C_{V/S} $  as the fraction of the incident gravitational energy flux of vector/scalar type, converted into electromagnetic energy flux, \cite{Gunter1980} shows that $ C_S\geq C_V $. The equality holds for $ Q=0 $ for which $ C_S=C_V=0 $. \emph{Hence, in RN, the scattered gravitational radiation due to a purely gravitational incoming wave will have less contribution from scalar perturbations compared to vector perturbations. }

\begin{figure}
	\centering
	\subfloat[][]{\includegraphics[width=0.4\linewidth, height=0.22\textheight]{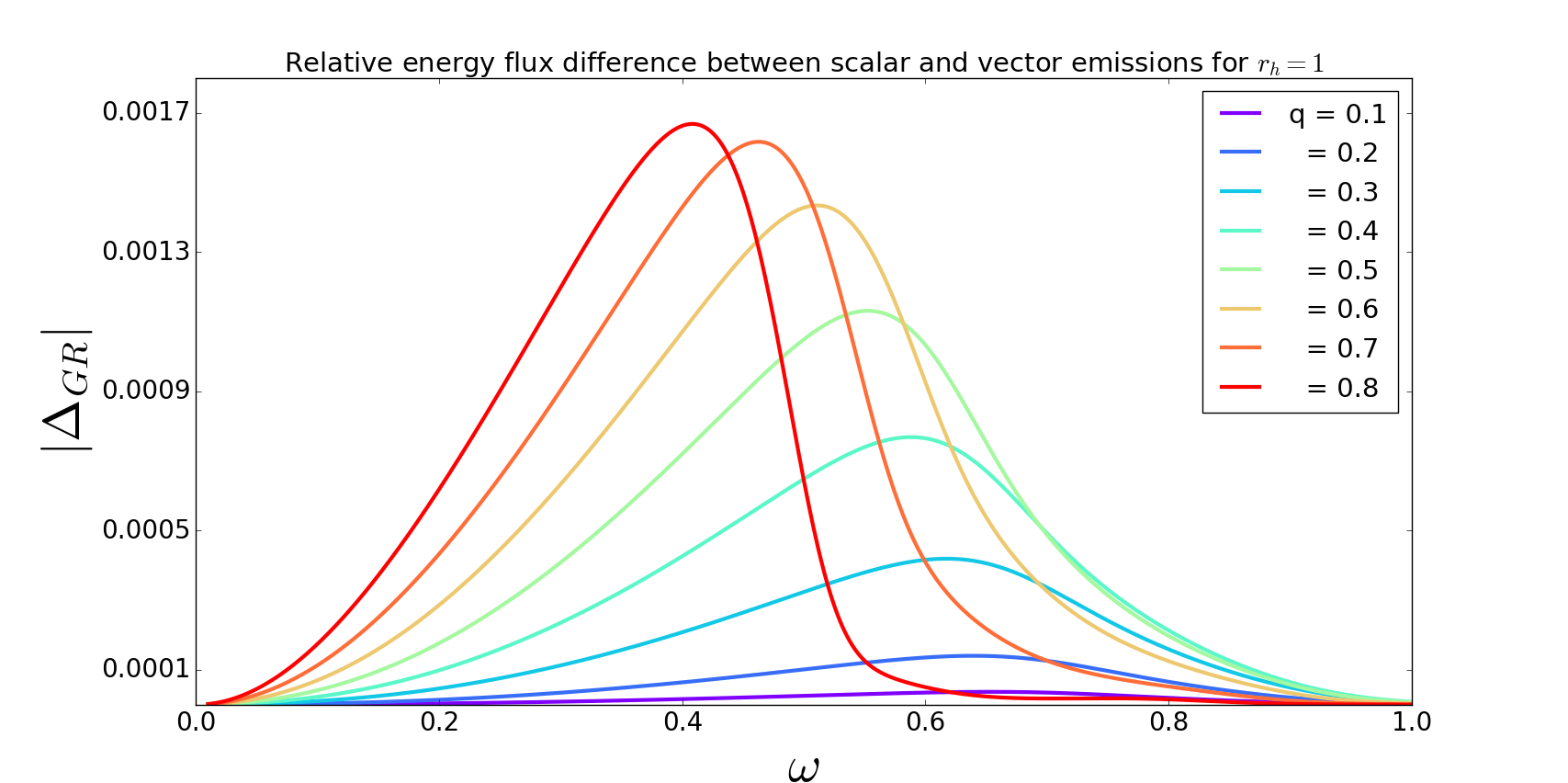}}\\
	\subfloat[][]{\includegraphics[width=0.4\linewidth, height=0.22\textheight]{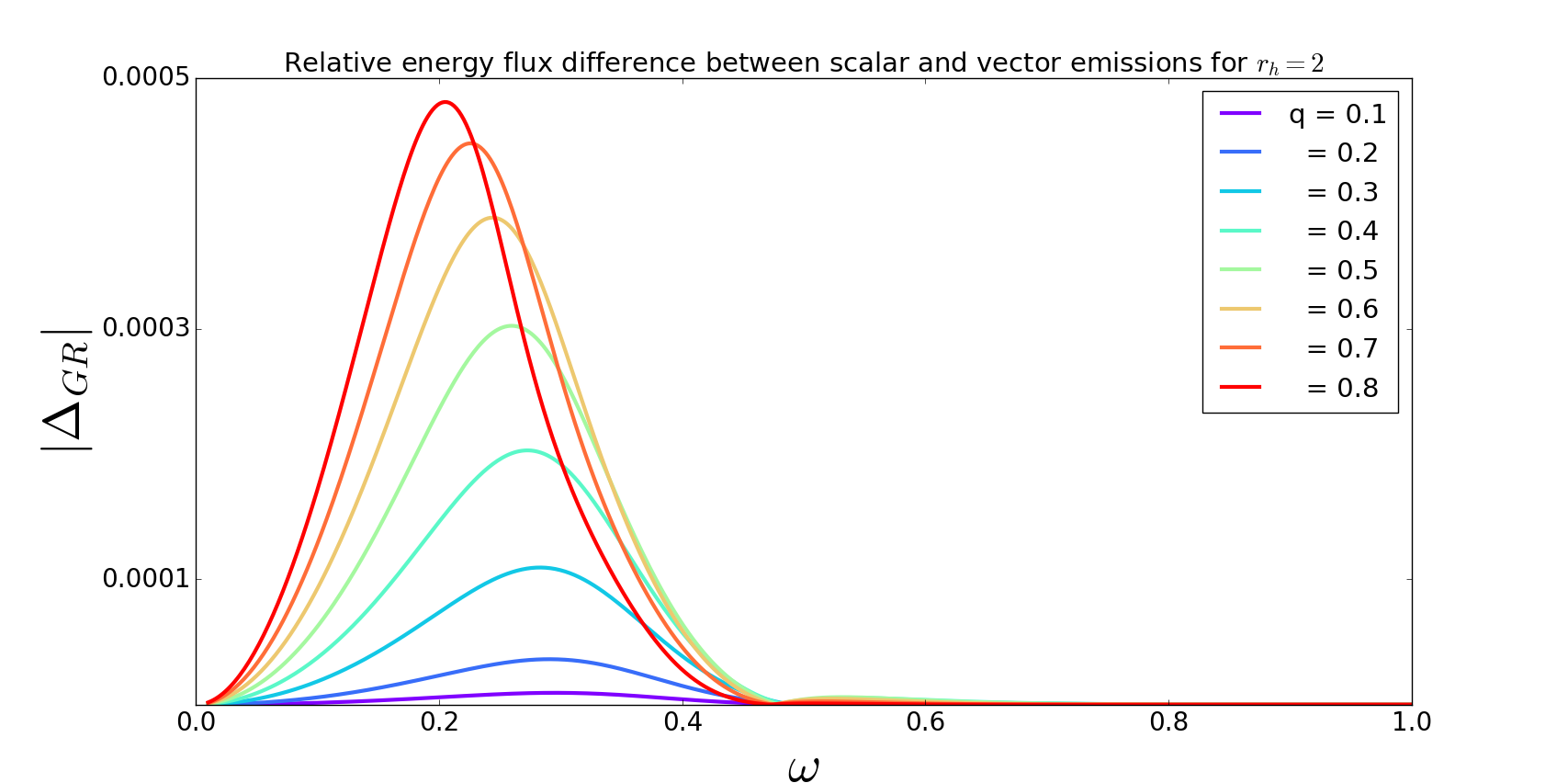}}\\
	\subfloat[][]{\includegraphics[width=0.4\linewidth, height=0.22\textheight]{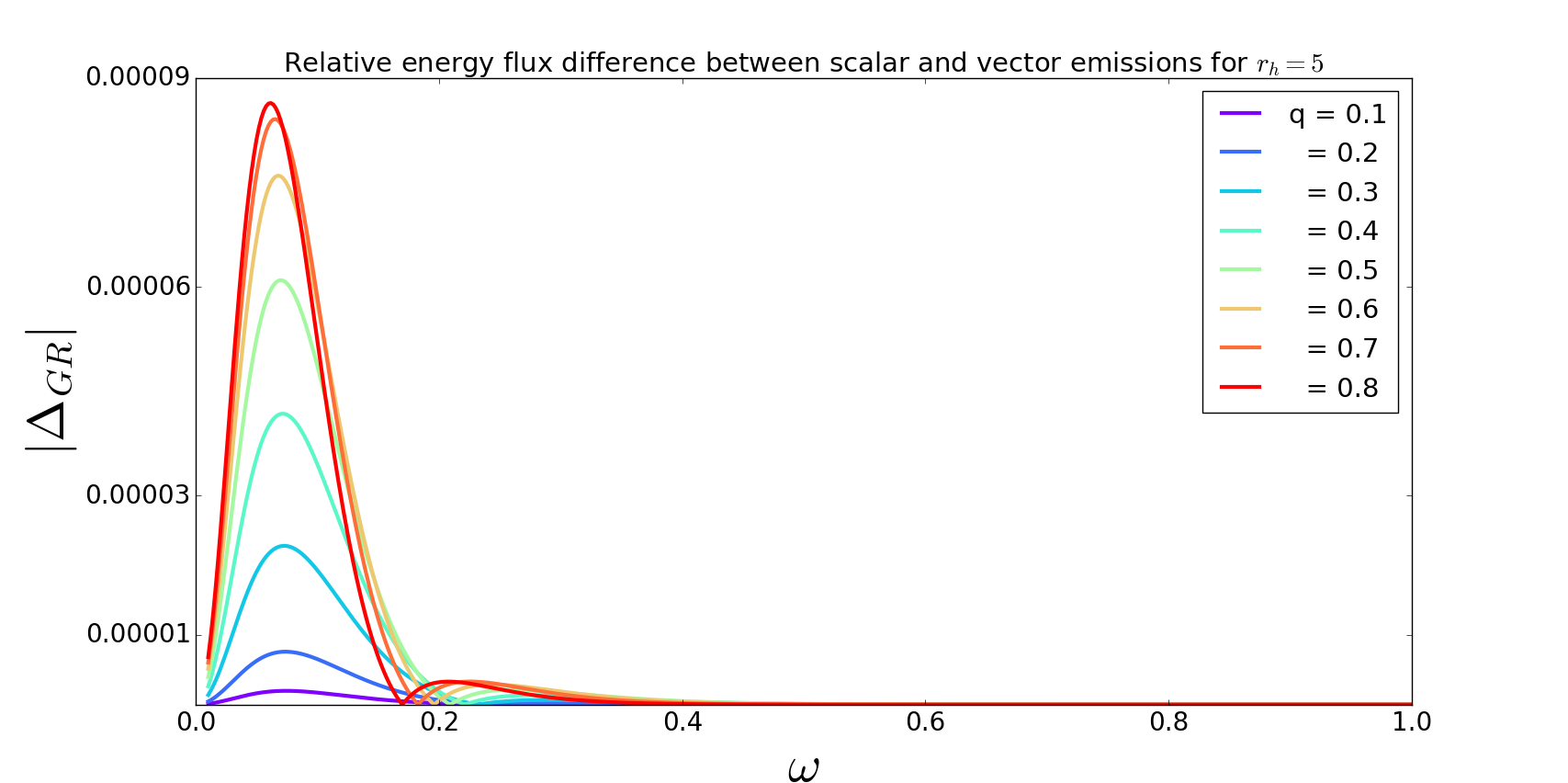}}
	\caption{Relative energy flux difference $ \Delta_{GR} $ for three different horizon radii. Increasing BH size leads to larger characteristic length scales for the space-time leading to the shift of the profiles towards low frequencies.}
		\label{fig:del_GR}
\end{figure}

\begin{figure}[h]
\centering
\includegraphics[width=0.6\linewidth, height=0.3\textheight]{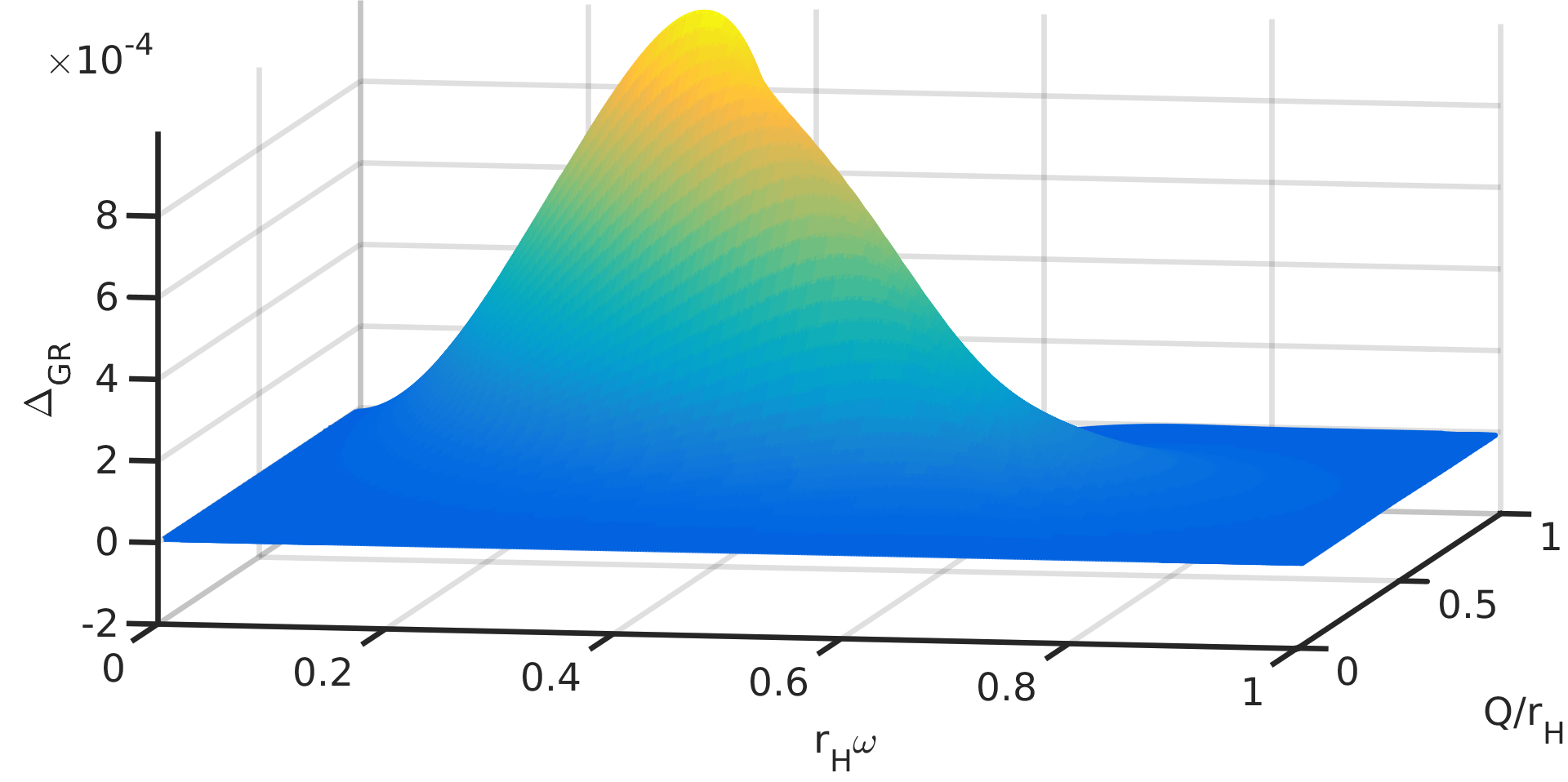}
\caption{Relative energy flux difference between scalar and vector emissions as a function of scaled charge q and dimensionless frequency $\tilde{\omega}$.}
\label{fig:En_Rel_RN}
\end{figure}

In the Schwarzschild case the scattered energy fluxes are equally distributed between the scalar and vector modes \cite{Chandrasekhar:1985kt,Bhattacharyya2017}. In RN spacetimes, the fraction of the gravitational flux radiated through the vector mode, that reaches an observer, compared to the net \textit{(gravitational+electromagnetic)} radiated vector mode flux is $ 1-C_V $. Similarly, the gravitational fraction of the net scalar mode flux is $ 1-C_S $. Thus, the relative difference in scattered gravitational energy fluxes between scalar and vector perturbations can be defined as:
\begin{eqnarray}\label{delGR}
\Delta_{GR}=\frac{(1-C_V)-(1-C_S)}{1-C_V}=\frac{C_S-C_V}{1-C_V} \, .
\end{eqnarray}
Appendix~\ref{conv} contains the details of the calculation of the conversion factors. \ref{fig:del_GR} describes the behavior of $\Delta_{GR}$ with respect to the scaled charge ($q$) and QNM frequency ($\omega$), whereas \ref{fig:En_Rel_RN} shows the behavior with respect to $ q $ and dimensionless frequency $ \tilde{\omega} $. From \ref{fig:del_GR} and \ref{fig:En_Rel_RN} it is clear from the mode dependence of $ C_{V/S} $ that $ \Delta_{GR} $ between the two massless modes give a measure of the black hole charge. Importantly, for a non-spinning hole, an energetic difference would imply that the black hole possesses charge. We will use this result in Sec. \ref{quantfR} to obtain a quantifying tool between GR and $f(R)$.
\pagebreak
%
\section{Perturbations in curved Ricci flat space-times for $ f(R) $ theories} \label{fR}

\subsection{Linear order perturbation equations}

The Lagrangian density of an $ f(R) $ theory can be represented by a polynomial of the form $ f(R)=R+\alpha R^2+\beta R^3+\mathcal{O}(R^4) $. The metric perturbations $h_{\mu\nu}$  about a background space-time and a Ricci perturbation about $\overline{R} =0 $ (which is a solution of polynomial $ f(R) $ theories) can help us truncate terms of powers higher than $ 2 $ \cite{Bhattacharyya2017}. Thus, we retain only the quadratic term
\begin{eqnarray}\label{fRaction}
\mathcal{L}=R+\alpha R^2.
\end{eqnarray}
Using the bounds found from E\"{o}t-Wash and fifth force tests \cite{Berry2011}, in this work, we set $ \alpha=10^{-9}  \mbox{m}^2 $. 
The equations of motion corresponding to the above action (\ref{fRaction}) are 
\begin{eqnarray}
G_{\mu\nu} &=&  T^{eff}_{\mu\nu} + \kappa^2T_{\mu\nu} \label{fieldeq} \\
T^{eff}_{\mu\nu} &=&  2 \alpha\left[R_{;\mu\nu}-g_{\mu\nu}\Box R+\frac{1}{4}g_{\mu\nu}R^{2}-RR_{\mu\nu}\right],\label{Teff} \nonumber 
\end{eqnarray}
where the higher derivative terms arising due to $ \alpha R^2 $ term in $ \mathcal{L} $ are referred to as effective energy-momentum tensor and $ T_{\mu\nu} $ was defined in (\ref{emem}). As shown in Ref. \cite{Bhattacharyya2017}, perturbations of $ T^{eff}_{\mu\nu} $ can be mapped to parameters (density, pressure, flux, anisotropic pressure) of a space-time fluid. This fluid acts as an extra massive scalar degree of freedom \cite{Berry2011}.

Perturbation about a curved background space-time $ \overline{g}_{\mu\nu} $  is given by Eq.~(\ref{eq:metricpert}). Electromagnetic field equations in curved space-times in Lorentz gauge are given by
\begin{eqnarray}
\Box A_\nu - R_{\mu\nu} A^\mu &=& 0 \label{cureom}
\end{eqnarray}
a perturbation of the 4-potential is given by
\begin{eqnarray}
A_\mu &=& \bar{A}_\mu+A^{(1)}_\mu
\end{eqnarray}
Dynamics of the extra degree of freedom is contained in $ h_{\mu\nu} $. In order to extract the extra field out off $ h_{\mu\nu} $, the perturbation variable can be redefined as \cite{Berry2011}
\begin{eqnarray}
\psi_{\mu\nu} &=& h_{\mu\nu} - \overline{g}_{\mu\nu}\left(\frac{h}{2} + 2\alpha R^{(1)} \right)  \label{redef}
\end{eqnarray}
where $ \psi_{\mu\nu} $ is a purely spin-2 field describing the two polarization modes and $ R^{(1)} = \overline{g}^{\alpha\beta}  R^{(1)}_{\alpha\beta} $ is the extra scalar degree of freedom. Using the Lorentz gauge condition~\cite{Berry2011,Isaacson1968b} for the gravitational and electromagnetic fields
\begin{eqnarray}
\psi^{\mu\nu}_{;\mu} &=& 0, \label{gauge}\\
A^{(1)\mu}\,_{;\mu} &=& 0 \, , \label{emgauge}
\end{eqnarray}
linearization of Eq.~(\ref{fieldeq}) and (\ref{cureom}) leads to
\begin{eqnarray}
\Box \psi_{\mu\nu} + 2 \overline{R}_{\alpha\mu\beta\nu}\psi^{\alpha\beta} &=& \kappa^2\left(\mathcal{U}_{\mu\nu}+\mathcal{T}_{\mu\nu}\right) \label{perteq} \\
\Box A_\nu^{(1)} &=& \mathcal{V}_{\nu} + 2\kappa^2\bar{T}_{\mu\nu}A^{(1)\mu} \label{perteqem}
\end{eqnarray}
where
\begin{eqnarray}
\mathcal{U}_{\alpha\beta} &=& 2\psi^{\mu\nu}\bar{F}_{\alpha\mu}\bar{F}_{\beta\nu}-\bar{g}_{\alpha\beta}\psi^{\mu\nu}\bar{F}_{\nu\rho}\bar{F}^{\rho}_{\mu}-2\bar{F}^{\nu}_{\mu}\bar{F}^{\mu}_{\left(\alpha\right.}\psi_{\left.\beta\right)\nu} \nonumber \\
\\
\mathcal{T}_{\alpha\beta} &=&  -2F^{(1)}_{\alpha\mu}\bar{F}^{\mu}_\beta - 2\bar{F}_{\alpha\mu}F^{(1)\mu}_\beta + \bar{g}_{\alpha\beta}\bar{F}.F^{(1)}\\
\mathcal{V}_\nu &=& 2\psi^{\alpha\beta}\bar{F}_{\alpha\nu;\beta}+\psi^{\beta;\alpha}_\nu \bar{F}_{\alpha\beta}
\end{eqnarray}
See Appendix (\ref{app:sourceterm}) for more details. The extra degree of freedom follows from the trace of the linearized equation (\ref{fieldeq})
\begin{eqnarray}
\Box R^{(1)} - \gamma^2 R^{(1)} &=& 0 \label{perteqex}
\end{eqnarray}
where $ \gamma=1/\sqrt{6\alpha}$. Eqs. (\ref{perteq}), (\ref{perteqem}), and (\ref{perteqex}) together describe the dynamics of perturbations in $f(R)$ theories.

\subsection{Energy-Momentum pseudotensor of perturbation} \label{empert}

As mentioned earlier, the perturbed black-holes respond by emitting gravitational waves. Isaacson \cite{Isaacson1968a,Isaacson1968b} was the first to quantify the energy-momentum density carried by a gravitational perturbation in general relativity by evaluating the energy-momentum pseudotensor. Energy-momentum pseudotensors have been studied previously in \cite{Aguirregabiria:1995qz,*Virbhadra:1990vs},\cite{Capozziello:2017xla}.
For $f(R)$ theories, the net field tensor of a perturbed Reissner-N\"{o}rdstrom space-time
\begin{equation}
\mathfrak{G}_{\mu\nu}\equiv G_{\mu\nu}-T^{eff}_{\mu\nu} - \kappa^2T_{\mu\nu}
\end{equation}
can be expanded in powers of $ \epsilon $ as follows
\begin{eqnarray}
\mathfrak{G}_{\mu\nu} &=& \bar{\mathfrak{G}}_{\mu\nu} + \epsilon \mathfrak{G}^{(1)}_{\mu\nu} + \epsilon^2 \mathfrak{G}^{(2)}_{\mu\nu} \, .
\end{eqnarray}
Following Isaacson \cite{Isaacson1968a}, the energy-momentum density carried by the gravitational and electromagnetic waves in $f(R)$ theories can be quantified as
\begin{eqnarray}
\bar{\mathfrak{G}}_{\mu\nu} &=& -\epsilon^2\langle \mathfrak{G}^{(2)}_{\mu\nu} 
\rangle  \equiv  \kappa^2 \, t_{\mu\nu} \\
t_{\mu\nu} &=&-\frac{\epsilon^2}{\kappa^2}\langle \mathfrak{G}^{(2)}_{\mu\nu} \rangle 
\end{eqnarray}
where $ \langle ... \rangle $ denotes averaging over small wavelengths~ \cite{Isaacson1968a}. The averaging process is done over a region where the change in background curvature is negligible compared to the change in perturbation variables - yielding a gauge invariant measure of the energy-momentum of perturbation.  (For details, see Appendix \ref{App:EMPseudotensor}.) The averaging procedure retains the gauge invariance of gravitational waves. For a Ricci flat space-time, the energy-momentum tensor of perturbation is given by
{\small
\begin{eqnarray} \label{pertem}
&  & t_{\mu\nu} = t_{\mu\nu}^{\textbf{I}} +  t_{\mu\nu}^{\textbf{II}} +  t_{\mu\nu}^{\textbf{III}} + t^{\textbf{IV}}_{\mu\nu} + \kappa^2\langle\mathcal{P}_{\mu\nu} \rangle  \\
&  & t_{\mu\nu}^{\textbf{I}} = -\frac{1}{4 \kappa^2} \bigg  \langle 
\psi^{\alpha\beta}_{;\mu}\psi_{\alpha\beta;\nu} \bigg \rangle \, ;
\quad  t_{\mu\nu}^{\textbf{II}} =  \frac{\alpha}{6 \kappa^2} \overline{g}_{\mu\nu}
\bigg  \langle \left(R^{(1)} \right)^2 \bigg \rangle \nonumber \\
& & t_{\mu\nu}^{\textbf{III}} = \frac{18 \alpha^2}{\kappa^2} \bigg  \langle 
 R^{(1)}_{;\mu} R^{(1)}_{;\nu} \bigg \rangle \, ; \quad t_{\mu\nu}^{\textbf{IV}} = 2\kappa^2 \bigg \langle A^{(1)\alpha}_{;\mu} A^{(1)}_{\alpha;\nu} \bigg \rangle \, ,
\end{eqnarray}
}
where the detailed calculations and the form of $ \langle \mathcal{P}_{\mu\nu} \rangle $ have been given in Appendix \ref{secav}. Eq.~(\ref{pertem}) allows us to draw the following important conclusions which is the first result of this paper. First, the energy density associated with the perturbation is $ t_{00} $. $ t_{00}^{\textbf{III}}$ is quadratic in $\alpha^2$ while $ t_{00}^{\textbf{II}}$ is linear in $\alpha$. Second, in the case of flat space-time, $ t_{00}^{\textbf{II}}$ vanish while $ t_{00}^{\textbf{I}} $ and $ t_{00}^{\textbf{III}} $ are non-zero. This provides an interesting prospective that the leading order corrections to general relativity go as $\alpha$ in curved space-time while as $\alpha^2$ in the flat space-time. Third,  $t_{00}^{\textbf{II}} $ vanishes at the horizon as $\overline{g}_{00}\rightarrow 0$ near the horizon. Thus, we have 
\begin{eqnarray}
t^{\textbf{II}}_{00} &\rightarrow& 0 ~~ \mbox{as} ~~r\rightarrow r_H \quad \left(\overline{g}_{00}\rightarrow 0\right)\\
&\rightarrow& 0 ~~ \mbox{as} ~~ r\rightarrow\infty \quad \left( R^{(1)} \rightarrow0\right)
\end{eqnarray}
The vanishing of $ t_{00}^{\textbf{II}} $ at the horizon and at infinity imply that the non-propagating contribution due to the modification of gravity exist around the black hole event horizon, which leaks energy of the order $ \alpha $ from the massless spin-2 mode to the massive extra degree of freedom.

\section{Quantifying tool for RN black-holes}\label{quantfR}

\subsection{Modified dynamics}
The perturbed effective energy momentum arising out of higher derivative terms of $ f(R) $ gravity, about $ \overline{R} =0 $, takes the form
\begin{eqnarray}
T^{(1),eff}_{\mu\nu} &=& 2\alpha\left(R^{(1)}_{\mu\nu} - \bar{g}_{\mu\nu}\Box R^{(1)} \right)
\end{eqnarray}
where we have ignored terms of $ \mathcal{O}\left(\alpha\kappa^2\right) $. Like the metric perturbation, this effective energy-momentum tensor can also be decomposed into scalar and vector spherical harmonics
\cite{Kodama2004b,Bhattacharyya2017}. Perturbations of the effective source term couples only to the scalar part of the massless perturbations leaving the vector part unchanged. As discussed in Sec. \ref{grdyn}, on linearizing (\ref{fieldeq}) and it's trace (\ref{perteqex}), by decomposing the metric and effective matter ($ T^{eff}_{\mu\nu} $) perturbation into scalar and vector part one obtains
\begin{eqnarray}
\frac{d^2\Phi^V_{\pm}}{dx^2}+\left(\tilde{\omega}^2-V^V_{\pm}\right)\Phi^V_{\pm} & =& 0 \label{fRpert1}\\
\frac{d^2\Phi^S_{\pm}}{dx^2}+\left(\tilde{\omega}^2-V^S_{\pm}\right)\Phi^S_{\pm} & =& S^{eff}_{\pm} \label{fRpert2} \\
\frac{d^2\Phi}{dx^2}+\left(\tilde{\omega}^2-\tilde{V}_{RW}\right)\Phi & =& 0 \label{massmod}
\end{eqnarray}
where $ x=r_*/{r_H} $, $ \tilde{\omega}=r_H\omega $. Perturbations of the extra mode involves perturbing the Ricci scalar and, as shown in \cite{Nzioki2014}, is given by $ \delta R=\Omega\textbf{S} $ and $ \Phi=r\Omega $ whose equation of motion was obtained from (\ref{perteqex}). It is important to note that the time dependence $ e^{i\omega t} $ was assumed for both the massless and the massive modes. The inhomogeneous effective source terms have the following forms \cite{*Kodama2004,Bhattacharyya2017}
\begin{eqnarray}
S^{eff}_{\pm}=c_{\pm}\tilde{\Phi}+d_{\pm}\frac{d\tilde{\Phi}}{dx} \label{effsauce}
\end{eqnarray}
where $ \tilde{\Phi}=\frac{4\alpha\Phi}{H} $, $ H(r)\equiv H=k^2-2+\frac{3(1+q^2)}{y}-\frac{4q^2}{y^2} $, and the coefficients $ (c_{\pm},d_{\pm}) $ depend on the parameters of the background space-time.

\subsection{An extra difference in energy densities}\label{delmod}

As seen in our earlier study for the Schwarzschild space-time \cite{Bhattacharyya2017}, modification to the Einstein-Hilbert action and its corresponding perturbed equations of motion \emph{only} modifies the scalar sector of the perturbation leaving the vector modes unchanged. In other words, the part of the perturbed energy density of the scalar sector leaks into this new massive mode, leading to a decreased energy density in the scalar modes. Due to the massive nature of the extra mode, excitations of the field around the black hole do not travel to asymptotic infinity and cannot be detected. However, the relative difference between the energy densities of scalar and vector modes can be used as an indirect probe to determine if such fields exist in nature.

As shown in Ref.~ \cite{Berry2011}, exciting the massive field in flat space is almost impossible, owing to the large frequency (and hence energy) cutoff (\ref{perteqex}) required by a gravitational perturbation to excite it. However, near a black hole that is not the case. To demonstrate this, we write down the exact form of the potential $ \tilde{V}_{RW} $
\begin{eqnarray}
\tilde{V}_{RW} & =& \left(1-\frac{1+q^2}{y}+\frac{q^2}{y^2}\right)\left(\frac{k^2}{y^2}+\frac{1+q^2}{y^3}-\frac{q^2}{y^4}+\frac{1}{6\tilde{\alpha}}\right), \nonumber \\
\end{eqnarray}
where $ \tilde{\alpha}=\frac{\alpha}{r_H^2} $. At large distances from the black hole $ (y\rightarrow\infty) $ $ \tilde{V}_{RW}\rightarrow\frac{1}{6\tilde{\alpha}} $, which gives back the flat space limit of the minimum frequency required to excite the massive field. However, for very small $ \alpha $, the potential can be approximated as
\begin{eqnarray}
\tilde{V}_{RW}\approx\left(1-\frac{1+q^2}{y}+\frac{q^2}{y^2}\right)\frac{r_H^2}{6\alpha}.
\end{eqnarray}
Thus near small black-holes the minimum frequency (and hence energy) required to excite the massive field is lesser than that of flat space.

\begin{figure}
	\centering
	\subfloat[][]{\includegraphics[width=10cm]{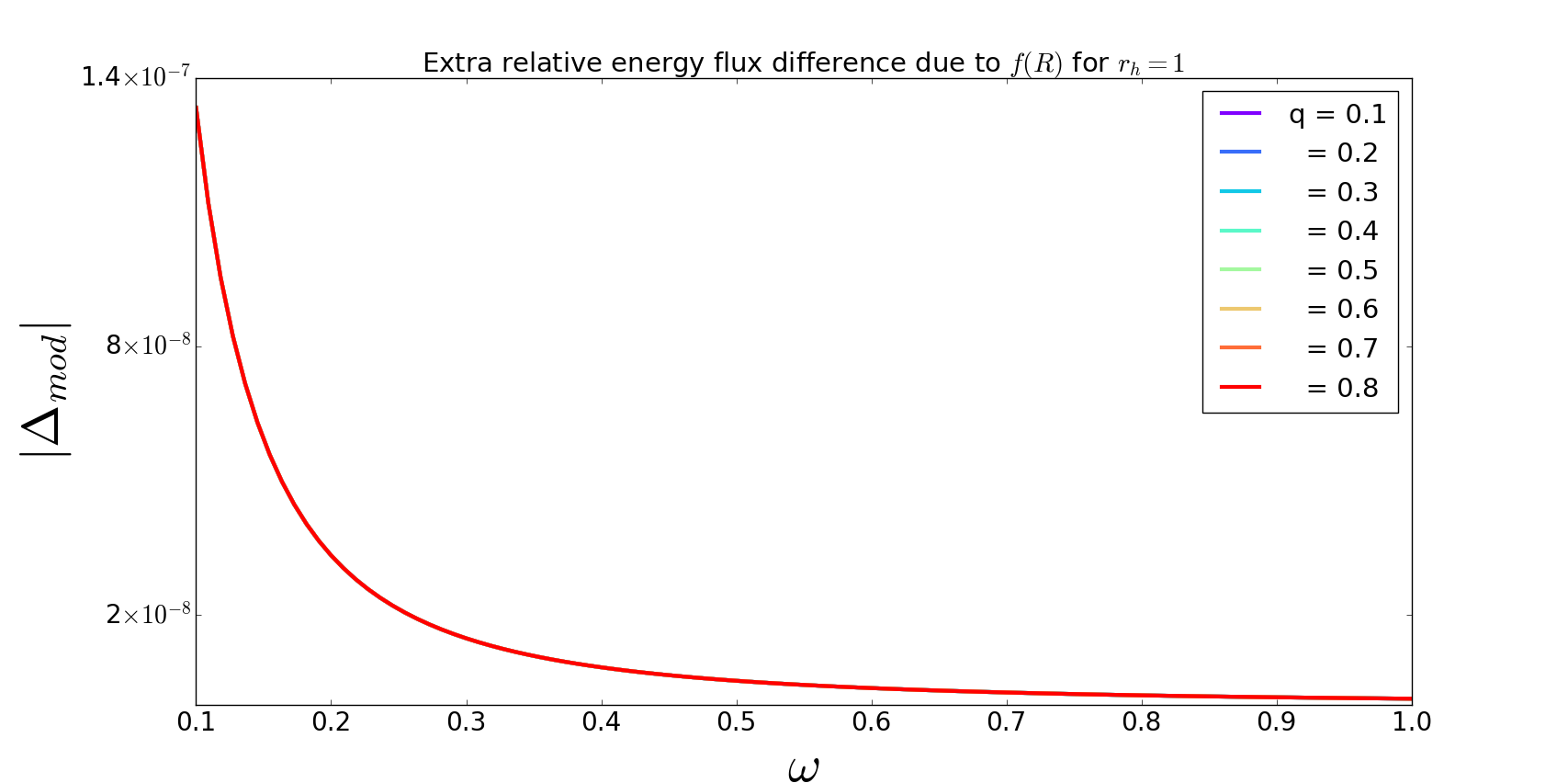}}\\
	\subfloat[][]{\includegraphics[width=10cm]{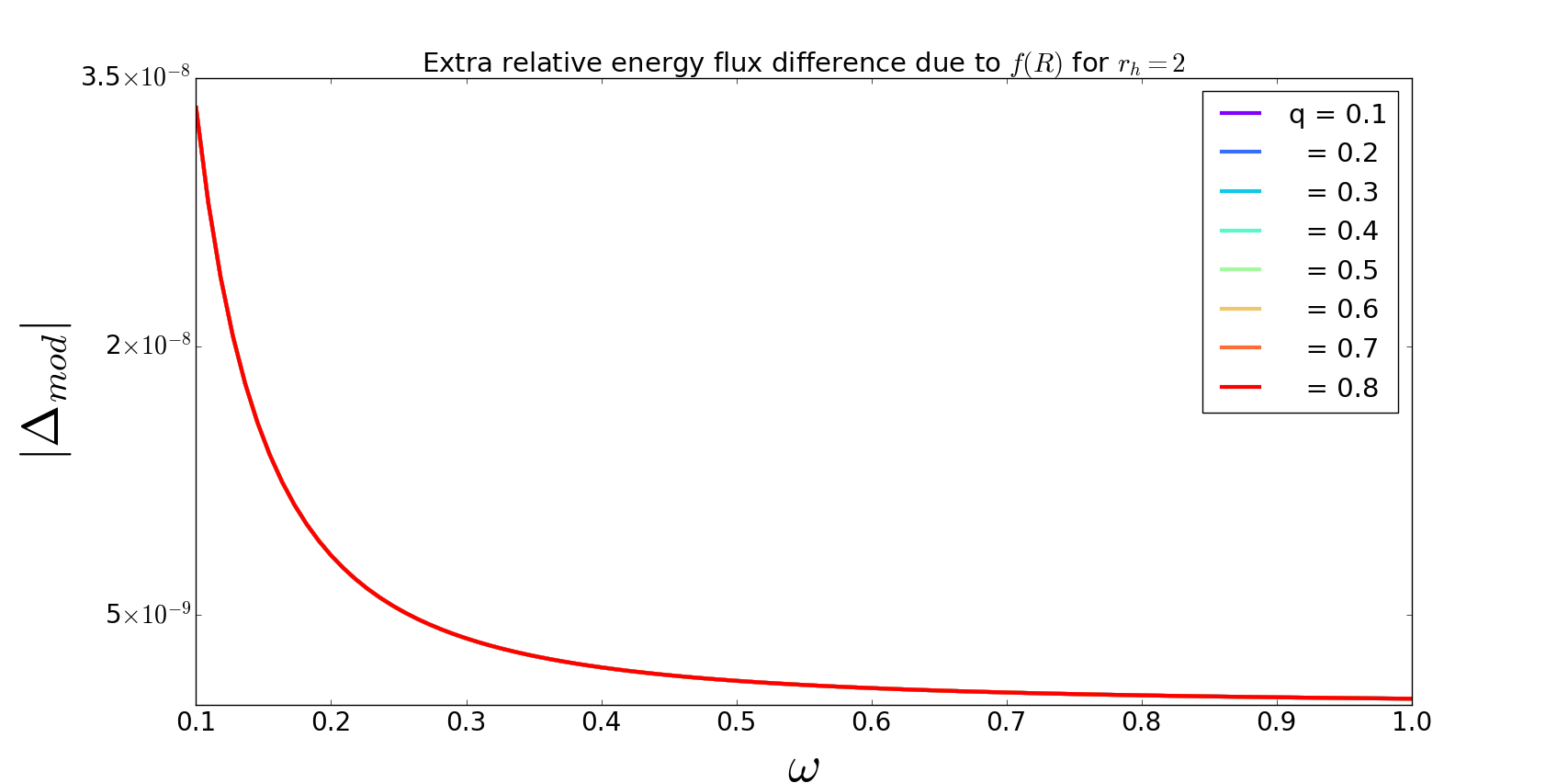}}\\
	\subfloat[][]{\includegraphics[width=10cm]{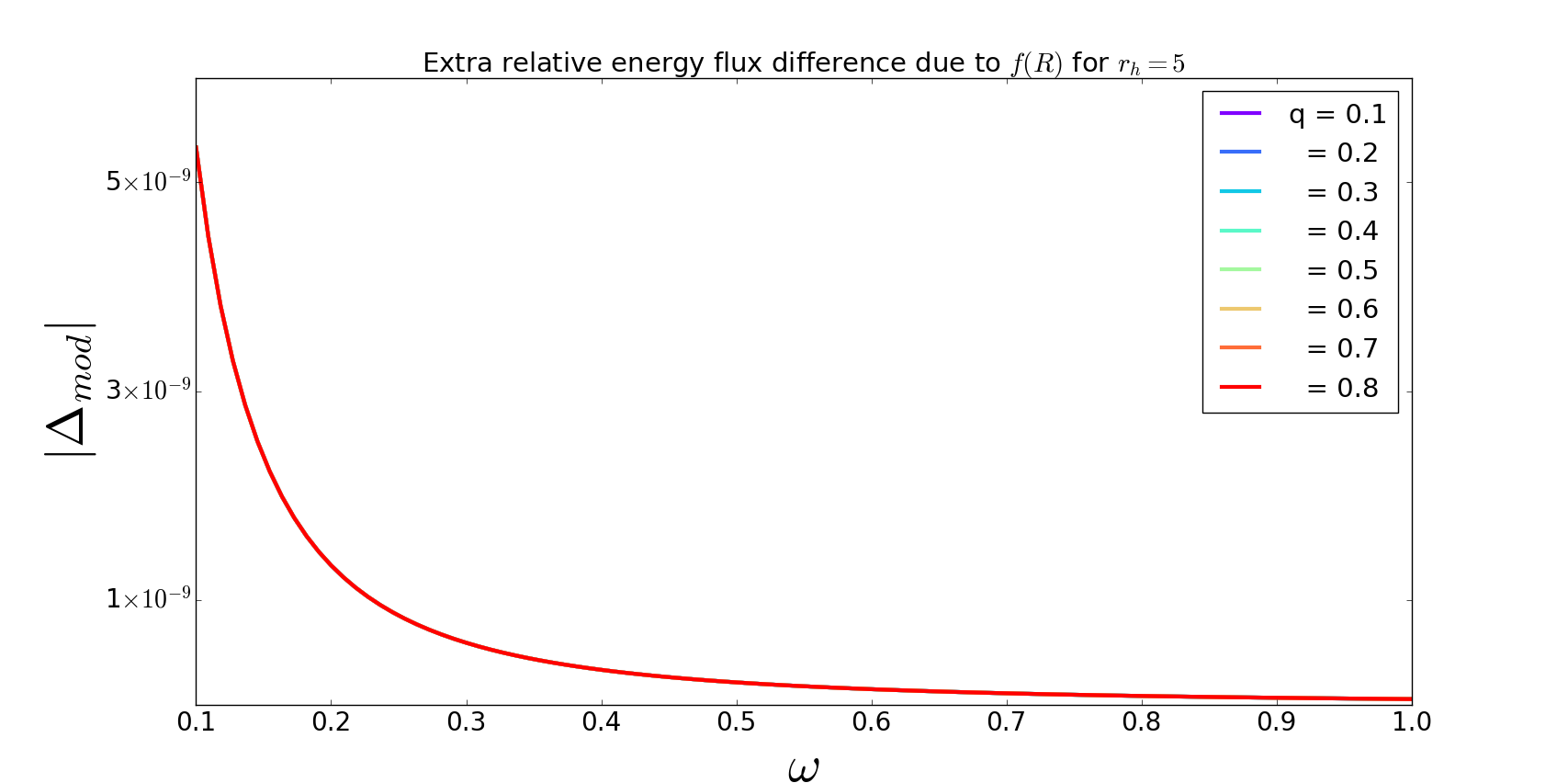}}
	\caption{Relative energy flux difference $ \Delta_{GR} $ for three different horizon radii. Increasing BH size leads to larger characteristic length scales for the space-time leading to the shift of the profiles towards low frequencies.}
	\label{fig:del_fR}
\end{figure}

\begin{figure}[h]
\centering
\includegraphics[width=0.7\linewidth, height=0.3\textheight]{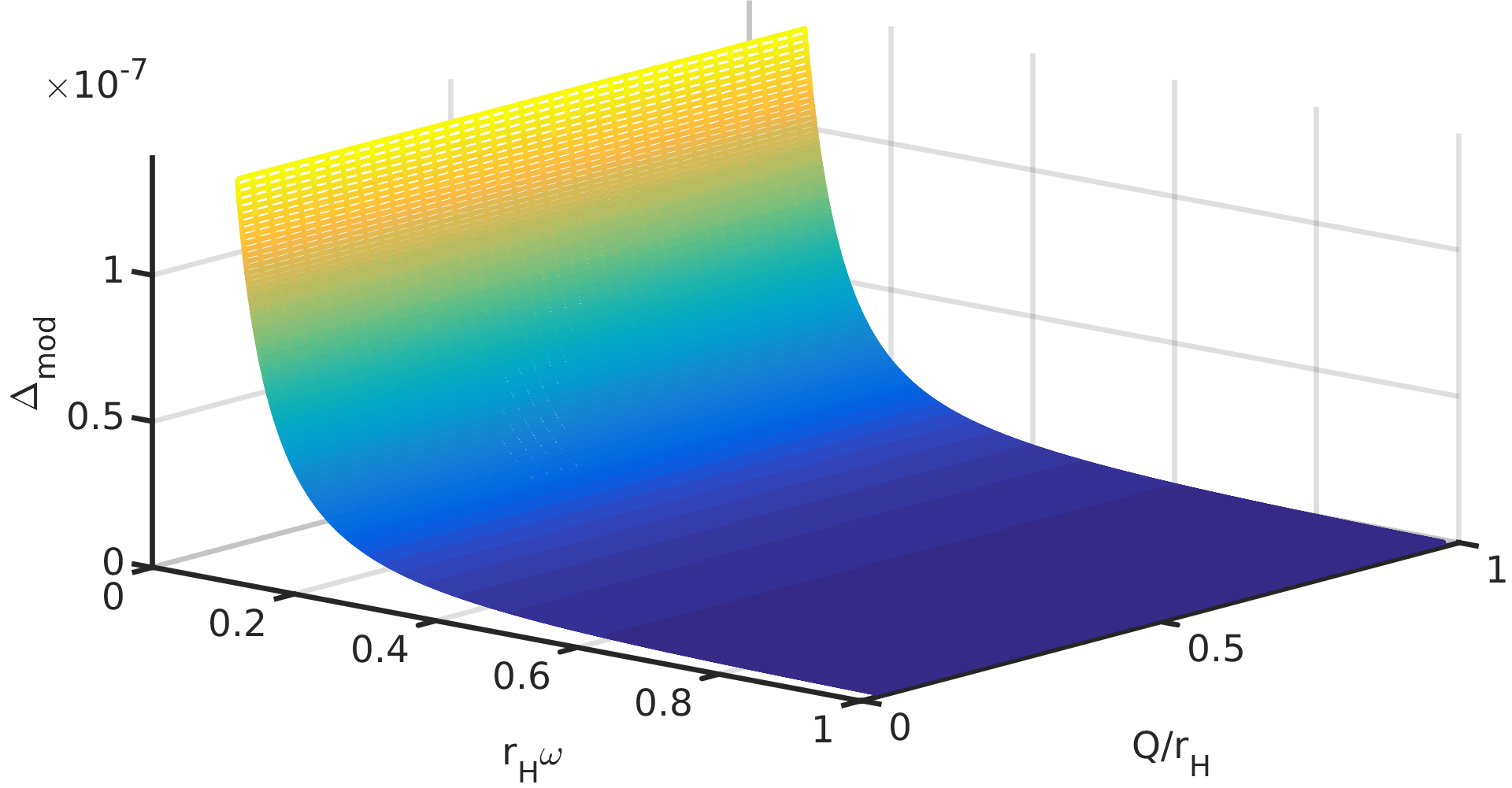}
\caption{$\Delta_{mod}$ as a function of the scaled charge $q$ and dimensionless frequency $\tilde{\omega}$.}
\label{fig:Rel_Energy_diff_RN}
\end{figure}

Eq.~(\ref{pertem}) contains the effective energy-momentum density of $ f(R) $ gravity due to a metric perturbation $ h_{\mu\nu} $ about the spherically symmetric space-time. We would like to highlight that the leading order corrections to the energy-momentum tensor in curved geometries near black-holes for $f(R)$ theories go as  $ \alpha $ in contrast to $ \alpha^2 $ in flat space. Ignoring the $ \alpha^2 $ contribution near black-holes, $ r_H $ scaled energetic deviation term is given by
\begin{eqnarray}
t^{\textbf{II}}_{\mu\nu} &=& \langle \frac{\alpha}{6r_H^4}\bar{g}_{\mu\nu}\left(\tilde{R}^{(1)}\right)^2 \rangle
\end{eqnarray}
where $ R^{(1)}=\frac{\tilde{R}^{(1)}}{r_H^2} $ and $ \tilde{R}^{(1)} $ is dimensionless. Thus, the energy density leaking into the massive mode from the massless modes is of the order $ \frac{\alpha}{6r_H^4} $ and the ratio of energetic difference between the scalar and vector modes can be defined as
\begin{eqnarray}
\Delta_{mod} &=& \left(1+\frac{1-C_V}{1-C_S}\right)\frac{t^{\textbf{II}}_{00}}{t^{\textbf{I}}_{00}}
\end{eqnarray}
where the term in the bracket indicates that the ratio is taken with respect to the radiated gravitational energy density in the scalar mode instead of the net \textit{(scalar+vector)} radiated gravitational energy density. Assuming the metric and rescaled Ricci perturbations are of the same order, and using Eq. (\ref{pertem}),  we get
\begin{eqnarray}
\Delta_{mod} &=& \frac{2}{3}\left(1+\frac{1-C_V}{1-C_S}\right)\frac{\alpha}{\tilde{\omega}^2r_H^2} \, . \label{delfR}
\end{eqnarray}
The above parameter is the difference in energy density between the vector and scalar perturbations for $f(R)$ theories of gravity.  This helps us to quantify a deviation from general relativity in the following manner:

\begin{itemize}
    \item Measure of the charge of a black hole obtained in general relativity from energetic difference of the massless modes and from the real \& imaginary parts of the quasinormal frequencies are equal.
    \item In $ f(R) $ theories, the extra massive field will lead to further energetic difference of the massless modes. Thus a measure of charge obtained from the energetic difference will be different from the one obtained from the quasinormal frequencies which only brings information about the three parameters mass, charge, and angular momentum - owing to \textit{no-hair} like theorems holding for $ f(R) $ theories \cite{10.1088/1361-6382/aa8e2e} and for Lovelock theories \cite{Skakala:2014gca}.
    \item The difference in the measured charge (the magnitude of difference depending on $ \alpha $) from the two methods should in principle allow us to detect and constraint any deviations from general relativity. Note that expression (\ref{delfR}) implies that deviation from GR is better detectable $ \left(\sim\frac{1}{r_H^2}\right) $ from the ringdown of smaller black-holes.
\end{itemize}
\ref{fig:del_fR} are plots of $ \Delta_{mod} $ versus charge $q$ and QNM frequency $\omega$ for $ \ell=2 $ and for three different horizon radii $ r_H $, whereas \ref{fig:Rel_Energy_diff_RN} is a 3-D plot of the same with respect to the scaled charge $ q $ and dimensionless frequency $ \tilde{{\omega}} $. Note that the behavior of $ \Delta_{mod} $ is dominated by the rescaled frequency ($ \frac{1}{\tilde{\omega}^2} $) and has a weak dependence on $ Q $ dependence through $ C_{V/S} $.

\section{Conclusions and Discussions}
As we have shown, the energetic equality between scalar and vector type gravitational perturbation is broken in the presence of charge in a black-hole. Hence, scattered gravitational radiation of the scalar and vector types carry different energies with them to asymptotic infinity - the extent of the difference depends on the magnitude of the charge. In other words, scalar type perturbations carry less energy to an observer at infinity compared to the vector type. This is because of the coupled nature of gravitational and electromagnetic perturbations in a Reissner-Nördstrom space-time and the fact that the degree of coupling depends on the perturbation type. 

A dimensionless parameter estimating the relative difference of energy flux between the scalar and vector modes was approximated using a parametrized P\"{o}schl-Teller potential which mimics the effective scattering potential of the two massless modes. A relationship between the spectrum of the relative difference between energy fluxes and the horizon radius scaled black hole charge was found, using which the charge of a spherically symmetric black hole in general relativity can be obtained from observations. We would like to mention to the readers that we have used P\"{o}schl-Teller potential as a preliminary estimate to obtain the reflection coefficients. Using higher-order WKB method and other semi-numerical methods it is possible to obtain correct estimates \cite{Kokkotas:1999bd,Berti:2009kk,0264-9381-16-12-201,Konoplya:2011qq}. We plan to obtain a more accurate numerical estimate in a future publication.

On a modification to gravity like $ f(R) $, there comes an
intrinsic extra massive degree of freedom in addition to
the two massless ones. The massive mode
couples itself only to the scalar type. We have shown explicitly that
this massive mode can be excited relatively easily around a
black-hole as compared to flat space. Hence, as seen in the 
RHS of (\ref{fRpert2}), the scattered
gravitational wave will have leakage of energy to the
massive mode owing to the ease of its excitation and a
coupling to the scalar type perturbation.  

Using Isaacson's prescription of calculating the energy-momentum pseudo-tensor of black-hole perturbations, we have obtained an accurate and robust method of estimating the difference in the radiated energies of the scalar and vector modes due to modifications to general relativity. The difference in the scattered energies of the massless modes for a charged black-hole will then have contributions from both the charge and the presence of the massive mode - which will result in the measure of charge using energetic difference (following \ref{fig:En_Rel_RN}) to be larger than the actual charge. Using the real and imaginary parts of the quasi-normal frequencies the charge of the black-hole can be obtained. Although astrophysical black-holes are expected to be charge-neutral, the time-scales involved in charge neutralization is usually longer than the time-scale in which black-hole forms in an NS-NS merger. Thus, the presence of charge after the merger can also cause a difference in radiated energies of the massless modes. It is then possible that the charge of a black-hole is so small that $ \Delta_{GR} $ becomes comparable to $ \Delta_{mod} $. But even in the tiny charge case, there will always be a difference in the two measurements (frequency and energetics) of the black-hole charge for a modification to gravity in the form of an intrinsic extra degree of freedom. This difference, if any, will constrain deviations from general relativity. However, it is necessary to note that the estimation of charge from the real and imaginary parts of the Quasinormal frequencies involve inclusion of the charge parameter in the template waveforms that are matched with the data to estimate all the parameters of the merging phenomenon. Currently, because of the immense computing power already required, templates taking charge into account are absent. However, better sensitivity and efficient algorithms of detection will help in expanding the parameter space for charge as well as modified gravity theories.

Gravitational radiation detected at asymptotic infinity
by a freely falling observer are characterized by two polarization amplitudes $ h_+ $ and $ h_\times $, related
to how distances between freely falling particles change
as a gravitational wave passes through them. The vector/scalar amplitudes are related to the polarization amplitudes in the following manner \cite{Bhattacharyya2017}
\begin{eqnarray}
\frac{d\Psi_S}{d\Omega} &=& h_+\textbf{S}_{\theta\theta} + \sin^2\theta h_\times\textbf{S}_{\theta\phi} \label{pol1}\\
\frac{d\Psi_V}{d\Omega} &=& h_+\textbf{V}_{\theta\theta} + \sin^2\theta h_\times\textbf{V}_{\theta\phi} \label{pol2}
\end{eqnarray}
where $ \Psi_{V/S} = \frac{\Phi_{V/S}}{r} $, $ \textbf{S}_{AB} $ \& $ \textbf{V}_{AB} $ are scalar and vector type harmonic tensors respectively \cite{Kodama2000}, and $ \Omega $ is the solid angle. Details of the above result are given in Appendix \ref{asympt}. The LHS of (\ref{pol1}) and (\ref{pol2}) quantify the flux of scalar and vector amplitudes that reach an observer at flat asymptotic infinity. In \cite{Abbott:2017oio}, individual intensities of the polarization of the $ h_{+/\times} $ modes were found from three detector observations. But the current resolution of the ringdown region is still poor, which is required to obtain charge and constrain deviations. In future detections, possibly in space based detectors like LISA, improved sensitivity would lead to a better resolution of the ringdown region of the gravitational waves, especially for NS-NS mergers. For such accurate observations it is then possible, in principle, to obtain the flux content of the scalar and vector modes from the polarization intensities of the ringdown stage.

Although in this work, we have focused on spherically symmetric space-times with the matter, the analysis in Sec.  \ref{empert} holds for arbitrary curved space-times, including Kerr and Kerr-Newmann. In the case of Kerr, last two terms in Eq.  (\ref{pertem}) vanishes. It is possible to obtain a measure
of the energy leakage from the massless to the massive
mode as like in Eq. (\ref{delfR}). However, owing to the reduction in the symmetry of Kerr compared to Schwarzschild or Reissner-Nördstrom space-times, the breaking up of a general perturbation
into scalar and vector types (which was possible under
rotations in $ \mathcal{S}^2 $ in a spherically symmetric space-time) becomes a non-trivial problem and has so far only been done in special cases like slow rotation approximation \cite{Pani:2012bp,*Pani:2012vp}. We hope to address the perturbation of Kerr black-holes in $ f(R) $ theories in a later paper.

The analysis in the paper has focused on a specific form of modified theories of gravity, and we plan to extend the analysis to other theories of gravity for spherically symmetric space-times. 

\section*{Acknowledgments}
The authors would like to thank C. P. L. Berry and A. Nagar for clarifications through email. SB is financially supported by the MHRD fellowship at IISER-TVM and would like to thank IIT Bombay for hospitality. The work is supported by DST-Max Planck-India Partner Group on Gravity and Cosmology.
\appendix

\section{Dynamics of gravitational and electromagnetic perturbations in general relativity} \label{grkod}
\subsection{Vector perturbations}
In 2+2 dimensions, the master variables $ \Phi_{V}^0 $ and $ \mathcal{A}_V $ for the background $ ds^2(g_{ab}) $
\begin{eqnarray}
ds^2(g_{ab}) & =& -\left(1-\frac{2M}{r}+\frac{Q^2}{r^2}\right)dt^2+\frac{dr^2}{\left(1-\frac{2M}{r}+\frac{Q^2}{r^2}\right)},\nonumber
\end{eqnarray}
satisfy the following coupled differential equations for source free (sources other than electromagnetic fields) space-times \cite{Kodama2004b}
\begin{eqnarray}
r^2D_a\left(\frac{1}{r^2}D^a\Omega\right)-\frac{k^2-1}{r^2}\Omega & =& \frac{2\sqrt{2}\kappa Q}{r^2}\mathcal{A}_V \label{decupV1}\\
\tilde{\Box}\mathcal{A}_V-\frac{1}{r^2}\left(k^2+1+\frac{4Q^2}{r^2}\right)\mathcal{A}_V & =& \frac{\sqrt{2}Q\left(k^2-1\right)}{\kappa r^4}\Omega \label{decupV2} \nonumber \\
\end{eqnarray}
where $ \Phi_V^0=\frac{\Omega}{r} $, $ \tilde{\Box}\equiv g^{ab}D_aD_b $, and $ k^2=\ell(\ell+1) $.

To decouple (\ref{decupV1}) and (\ref{decupV2}) two new variables $ \Phi_{\pm}^V $ are defined as
\begin{eqnarray}
\Phi_\pm^V & =& a_\pm^V\Phi^0_V+b_\pm^V\mathcal{A}_V \label{decuptrans} \\
\left(a_+^V,b_+^V\right) & \equiv& \left(\frac{Q(k^2-1)}{3M+\Delta},\frac{\kappa}{\sqrt{2}}\right) \\
\left(a_-^V,b_-^V\right) & \equiv& \left(1,\frac{-2\sqrt{2}\kappa Q}{3M+\Delta}\right)
\end{eqnarray}
where $ \Delta=\sqrt{9M^2+4(k^2-1)Q^2} $.

Using (\ref{decuptrans}) on (\ref{decupV1}) and (\ref{decupV2}), transforming to tortoise coordinates, and scaling with respect to $ r_H $ one obtains (\ref{GRpert1})
\begin{eqnarray}
\frac{d^2\Phi^V_{\pm}}{dx^2}+\left(\tilde{\omega}^2-V^V_{\pm}\right)\Phi^V_{\pm} & =& 0
\end{eqnarray}
where the unscaled $ V_\pm^V $ is given by
\begin{eqnarray}
V_\pm^V=\frac{g}{r^2}\left(k^2+1+\frac{4Q^2}{r^2}+\frac{-3M\pm\Delta}{r}\right) \, .
\end{eqnarray}
Note that for $ r\rightarrow r_H $, $ g\rightarrow0 $ and $ V_\pm^V\rightarrow0 $, for $ r\rightarrow\infty $, $ V_\pm^V\rightarrow0 $, and $ V_\pm^V\geq0 $ for $ r_H\leq r<\infty $ - indicating the short ranged scattering nature of the potentials.
\subsection{Scalar perturbations}
The master variables $ \Phi^0_S $ and $ \mathcal{A}_S $ satisfy the following coupled system in 2+2 dimensions
\begin{eqnarray}
g\frac{d}{dr}\left(g\frac{d\Phi^0_S}{dr}\right)+\left(\omega^2-V_S\right)\Phi^0_S & =& 0 \label{decupS1} \\
\tilde{\Box}\mathcal{A}_S-\frac{1}{r^2}\left(k^2+\frac{8Q^2g}{r^2H}\right)\mathcal{A}_S & =& \frac{\sqrt{2}Q}{\kappa r^3}\left(\frac{2H^2-P_Z}{4H}\Phi^0_S\right.\nonumber\\
&&\left.+gr\frac{d\Phi^0_S}{dr}\right)  \label{decupS2}
\end{eqnarray}
where $ H=k^2-2+\frac{6M}{r}-\frac{4Q^2}{r^2} $, and $ P_Z={\frac {{8M}^{2}}{{r}^{2}}}+{\frac {2M}{r} \left( {-\frac {4{Q}^{2}}{{r}^{2}}}+6\,{k}^{2}-6 \right) }-{\frac { 8\left( {k}^{2}+1 \right){Q}^{2}}{{r}^{2}}}-4\,{k}^{2}+8 $.

Eqs. (\ref{decupS1}) and (\ref{decupS2}) can be decoupled by the following transformation
\begin{eqnarray}
\Phi_\pm^S & =& a_\pm^S\Phi^0_S+b_\pm^S\mathcal{A}_S \label{decuptransS}\\
\left(a_+^S,b_+^S\right) & \equiv& \left(\frac{ Q({k}^{2}r+3\,M+3\,\mu-2\,r)}{2r},\frac{3\left(M+\mu\right)\kappa}{\sqrt{2}}\right) \nonumber\\
\\
\left(a_-^S,b_-^S\right) & \equiv& \left(3(M+\mu)-\frac{4Q^2}{r},-\frac{4\sqrt{2}Q}{\kappa}\right)
\end{eqnarray}
where $ \mu=\sqrt{M^2+\frac{4}{9}(k^2-2)Q^2} $.

Using (\ref{decuptransS}) in (\ref{decupS1}) and (\ref{decupS2}), transforming to tortoise coordinates, and scaling with respect to $ r_H $ one obtains (\ref{GRpert2}):
\begin{eqnarray}
\frac{d^2\Phi^S_{\pm}}{dx^2}+\left(\tilde{\omega}^2-V^S_{\pm}\right)\Phi^S_{\pm} & =& 0
\end{eqnarray}
where $ V_\pm^S $ are scattering potentials similar to $ V_\pm^V $ and have the same reflection and transmission coefficients.
\section{Parametrized P\"{o}schl-Teller potential approach for calculating conversion factors $ C_{V/S} $} \label{conv}
\subsection{Decoupled nature of the Gravitational and Electromagnetic perturbations}
At the boundaries $ x\rightarrow\pm\infty $, $ \Phi_\pm^{V/S} $ is given by
\begin{eqnarray}
\Phi_\pm^{V/S} & \sim& e^{i\tilde{\omega}x}+\sqrt{R_\pm^{V/S}}e^{i\left(\delta_{\pm,(r)}^{V/S}-\tilde{\omega}x\right)} \qquad x\rightarrow\infty \\
\Phi_\pm^{V/S} & \sim& \sqrt{T_{\pm}^{V/S}}e^{i\left(\delta_{\pm,(t)}^{V/S}+\tilde{\omega}x\right)} \qquad x\rightarrow-\infty
\end{eqnarray}
where $ R_\pm^{V/S} $ \& $ T_{\pm}^{V/S} $ are the reflection and transmission coefficients associated with the potentials $ V_\pm^{V/S} $, while $ \delta_{\pm,(r/t)}^{V/S} $ are the changes in phase of the incoming wave due to reflection/transmission off of the potential barriers  $ V_\pm^{V/S} $ .

A purely gravitational wave incident on the black-hole space-time from $ x\rightarrow\infty $ is given by
\begin{eqnarray}
\Phi^0_{V/S,(i)} & \neq& 0 \\
\mathcal{A}_{V/S,(i)} & =& 0
\end{eqnarray}
where the reflected wave is given by \cite{Gunter1980}
\begin{eqnarray}
&&\left|\Phi^0_{V/S,(r)}\right| = \left|\Phi^0_{V/S,(i)}\right|\left[R_+^{V/S}\sin^2\epsilon+R_-^{V/S}\cos^2\epsilon\right.\nonumber\\
&&\left.+2\sqrt{R_+^{V/S}R_-^{V/S}}\cos\left(\delta_{+,(r)}^{S/V}-\delta_{-,(r)}^{V/S}\right)\right]^\frac{1}{2}\sin\epsilon\cos\epsilon\nonumber\\
\\
&&\left|\mathcal{A}_{V/S,(r)}\right| = \left|\Phi^0_{V/S,(i)}\right|\left[R_+^{V/S}+R_-^{V/S}\right.\nonumber\\
&&\left.-2\sqrt{R_+^{V/S}R_-^{V/S}}\cos\left(\delta_{+,(r)}^{S/V}-\delta_{-,(r)}^{V/S}\right)\right]^\frac{1}{2}\sin\epsilon\cos\epsilon \label{convkey} \nonumber \\
\end{eqnarray}
\begin{eqnarray}
\sin2\epsilon &=& \mp2\sqrt{\frac{-q_1q_2}{\left(q_1-q_2\right)^2}} \qquad S(-), V(+)\\
q_i &=& 3M+\left(-1\right)^{i-1}\sqrt{9M^2+4\left(k^2-2\right)Q^2} \qquad i=1,2 \nonumber\\
\end{eqnarray}
indicating that a purely gravitational wave on scattering off of the curvature of a Reissner-N\"{o}rdstrom space-time will have a minor electromagnetic component in the net scattered radiation. The factor multiplying $ \left|\Phi^0_{V/S,(i)}\right| $ in (\ref{convkey}) is called the conversion factor ($ C_{V/S} $) and was first calculated in \cite{Gunter1980}, and shown that $ C_S\geq C_V $.

\subsection{The P\"{o}schl-Teller method}
To calculate $ C_{V/S} $ we need to calculate the absolute ($ \sqrt{R_\pm^{V/S}} $) and phase ($ \delta_{\pm,(r)}^{V/S} $) parts of the reflection amplitude of the form $ \sqrt{R_\pm^{V/S}}e^{i\delta_{\pm,(r)}^{V/S}} $ - which can be found by utilizing the method used in \cite{Ferrari1984}. In this scheme the potentials $ V_\pm^{V/S} $ are replaced by a properly parametrized P\"{o}schl-Teller potential which is of the form
\begin{eqnarray}
U_{PT}(x) & =& \frac{U_0}{\cosh^2\left[\beta(x-x_0)\right]}
\end{eqnarray}
where $ U_0=U_{PT}(x_0) $ is the maximum value of the potential and $ \beta=-\sqrt{\frac{1}{2U_0}\left.\frac{d^2U_{PT}}{dx^2}\right|_{x=x_0}} $ is the curvature about the maximum. Reflection amplitude for this potential was found from \cite{Ferrari1984}
\begin{eqnarray}
R(\omega)=\frac{\Gamma\left(\frac{-i\omega}{\beta}\right)\Gamma\left(1+\chi+\frac{i\omega}{\beta}\right)\Gamma\left(-\chi+\frac{i\omega}{\beta}\right)}{\Gamma\left(\frac{i\omega}{\beta}\right)\Gamma\left(1+\chi\right)\Gamma\left(-\chi\right)}\label{refl}
\end{eqnarray}
where $ \Gamma(a) $ is the Gamma function and $ \chi=-\frac{1}{2}+\sqrt{\frac{1}{4}-\frac{U_0}{\beta}} $. Absolute and phase parts of (\ref{refl}) give the reflection coefficient and the phase change on scattering respectively.

\section{Wave propagation in curved space and Lorentz gauge} \label{secav}
\subsection{Linearized equations of motion}
First order perturbed quantities of interest are given by
\begin{eqnarray}
G^{(1)}_{\alpha\beta} &=& R^{(1)}_{\alpha\beta}-\frac{1}{2}\bar{g}_{\alpha\beta}R^{(1)}\\
T^{(1),eff}_{\alpha\beta} &=& 2\alpha\left(R^{(1)}_{\mu\nu} - \bar{g}_{\mu\nu}\Box R^{(1)} - 2R^{(1)}\bar{R}_{\mu\nu}\right) \\
T^{(1)}_{\alpha\beta} &=& F^{(1)}_{\alpha\mu}\bar{F}^{\mu}_\beta + \bar{F}_{\alpha\mu}F^{(1)\mu}_\beta -h^{\mu\nu}\bar{F}_{\alpha\mu}\bar{F}_{\beta\nu} - \frac{1}{4}\left[h_{\alpha\beta}\bar{F}^2  \right.\nonumber \\
& & \left. - \bar{g}_{\alpha \beta} h^{\rho \mu} \bar{F}_{\mu \nu} \bar{F}^{\nu}_{\rho} - \bar{g}_{\alpha \beta} h^{\rho \nu} \bar{F}_{\mu \nu} \bar{F}^{\mu}_{\rho} + 2\bar{g}_{\alpha\beta}\bar{F}.F^{(1)}\right] \nonumber \\
\end{eqnarray}
The linearized Ricci tensor is given by \cite{Isaacson1968b}
\begin{eqnarray}
R^{(1)}_{\mu\nu} &=& \left(h_{;\mu\nu} + \Box h_{\mu\nu} - h^\rho_{\alpha;\beta\rho} - h^\rho_{\beta;\alpha\rho}\right)
\end{eqnarray}
Redefining the perturbation to be
\begin{eqnarray}
\psi_{\mu\nu} &=& h_{\mu\nu} - \bar{g}_{\mu\nu}\left(\frac{h}{2} + 2\alpha R^{(1)}\right) \label{redefapp}
\end{eqnarray}
and using the commutation relation of covariant derivatives \cite{Misner:1974qy}, $ \mathfrak{G}^{(1)}_{\mu\nu}=0 $ gives
\begin{eqnarray}
&&\Box \psi_{\mu\nu} + 2 \overline{R}_{\alpha\mu\beta\nu}\psi^{\alpha\beta} = \kappa^2\left(\mathcal{U}_{\mu\nu}+\mathcal{T}_{\mu\nu}\right) \\
&&\mathcal{U}_{\alpha\beta} = 2\psi^{\mu\nu}\bar{F}_{\alpha\mu}\bar{F}_{\beta\nu}-\bar{g}_{\alpha\beta}\psi^{\mu\nu}\bar{F}_{\nu\rho}\bar{F}^{\rho}_{\mu}-2\bar{F}^{\nu}_{\mu}\bar{F}^{\mu}_{\left(\alpha\right.}\psi_{\left.\beta\right)\nu} \nonumber \\
\\
&&\mathcal{T}_{\alpha\beta} =  -2F^{(1)}_{\alpha\mu}\bar{F}^{\mu}_\beta - 2\bar{F}_{\alpha\mu}F^{(1)\mu}_\beta + \bar{g}_{\alpha\beta}\bar{F}.F^{(1)}
\end{eqnarray}
subject to the following gauge conditions
\begin{eqnarray}
\psi_{\mu\nu}\,^{;\mu} &=& 0 \label{ggauge1} \\
\psi &=& 0 \label{ggauge2}
\end{eqnarray}
Similarly, the curved space Maxwell equations in Lorentz gauge is
\begin{eqnarray}
\Box A_\mu - R_{\mu\nu} A^\nu &=& 0
\end{eqnarray}
The linearization of which then leads to
\begin{eqnarray}
\Box A_\nu^{(1)} &=& \mathcal{V}_{\nu} + 2\kappa^2\bar{T}_{\mu\nu}A^{(1)\mu} \\
\mathcal{V}_\nu &=& 2\psi^{\alpha\beta}\bar{F}_{\alpha\nu;\beta}+\psi^{\beta;\alpha}_\nu \bar{F}_{\alpha\beta}
\end{eqnarray}
subject to the gauge condition
\begin{eqnarray}
A^{(1)\mu}\,_{;\mu} &=& 0 \label{egauge}
\end{eqnarray}
\subsection{Derivation of the energy-momentum pseudo-tensor}
\label{App:EMPseudotensor}
The averaging over several wavelengths as introduced by \cite{Isaacson1968a} involves the following guidelines
\begin{itemize}
	\item Total derivative terms of the form $ \langle A_{\alpha..\beta;\mu} \rangle = 0 $.
	\item $ \langle A_{;\alpha}B_{;\beta} \rangle = -\langle A_{;\alpha\beta}B \rangle $, where $ A $ and $ B $ are indexed tensor objects.
	\item Averages of a product of independent fields are zero.
\end{itemize}
Thus, various second order quantities become
\begin{eqnarray}
G^{(2)}_{\alpha\beta} &=& R^{(2)}_{\alpha\beta} - h_{\alpha\beta}R^{(1)} + \bar{g}_{\alpha\beta}h^{\mu\nu}R^{(1)}_{\mu\nu} - \frac{1}{2} \bar{g}_{\alpha\beta}\gamma^{\mu\nu}R^{(2)}_{\mu\nu} \nonumber \\
\label{G2}\\
T^{(2)}_{\alpha\beta} &=& 2F^{(1)}_{\alpha\mu}F^{(1)\mu}_\beta - \frac{1}{4}\left(2\bar{g}_{\alpha\beta}F^{(1)}.F^{(1)}\right. \nonumber\\
&&\left.- h_{\alpha\beta}h^{\rho\mu}\bar{F}_{\mu \nu} \bar{F}^{\nu}_{\rho} - h_{\alpha\beta} h^{\rho \nu} \bar{F}_{\mu \nu} \bar{F}^{\mu}_{\rho}\right) \label{T2}\\
T^{(2),eff}_{\alpha\beta} &=& \alpha\left[4\delta\nabla_\alpha R^{(1)}_{;\beta} - 2h_{\alpha\beta}\Box R^{(1)} + 2\bar{g}_{\alpha\beta}h^{\mu\nu}R^{(1)}_{;\mu\nu} \right. \nonumber\\
&&\left.- \bar{g}_{\alpha\beta}\delta\nabla^\mu R^{(1)}_{;\mu} + \bar{g}_{\alpha\beta}\left(R^{(1)}\right)^2 - 4R^{(1)}R^{(1)}_{\alpha\beta}\right] \nonumber\label{Te2}\\
\end{eqnarray}
where
\begin{eqnarray}
R^{(1)}_{\mu\nu} &=& \frac{1}{2}\left(h_{;\mu\nu} + \Box h_{\mu\nu} - h^\rho_{\alpha;\beta\rho} - h^\rho_{\beta;\alpha\rho}\right) \\
R^{(2)}_{\mu\nu} &=& -\frac{1}{2} \left[\frac{1}{2} h^{\rho\tau}_{;\beta} h_{\rho\tau;\alpha} + h^{\rho\tau} \left(h_{\rho\tau;\alpha\beta} + h_{\alpha\beta;\tau\rho} \right.\right.\nonumber\\
&&\left. - h_{\tau\alpha;\beta\rho} - h_{\tau\beta;\alpha\rho}\right) + h^{\tau;\rho}_\beta \left(h_{\tau\alpha;\rho} - h_{\rho\alpha;\tau}\right) \nonumber \\
&& \left. - \left(h^{\rho\tau}_{;\rho} - \frac{1}{2}h^{;\tau}\right) \left(h_{\tau\alpha;\beta} + h_{\tau\beta;\alpha} - h_{\alpha\beta;\tau}\right)\right] \nonumber \\
\\
F^{(1)}_{\mu\nu} &=& \partial_\mu A^{(1)}_\nu - \partial_\nu A^{(1)}_\mu
\end{eqnarray}
Using (\ref{redefapp}), (\ref{ggauge1}), (\ref{ggauge2}), (\ref{egauge}), and the commutation relation of the covariant derivatives \cite{Misner:1974qy} on (\ref{G2})-(\ref{Te2}) one obtains after ignoring $ \mathcal{O}(\kappa^4) $ terms
\begin{eqnarray}
\langle \mathfrak{G}^{(2)}_{\alpha\beta} \rangle &=& \frac{1}{4}\langle \psi^{\rho\tau}_{;\alpha}\psi_{\rho\tau;\beta} \rangle -\frac{1}{6}\alpha\bar{g}_{\alpha\beta}\langle\left(R^{(1)}\right)^2\rangle -18\alpha^2\langle R^{(1)}_{:\alpha}R^{(1)}_{;\beta} \rangle \nonumber \\
&&-2\kappa^2\langle A^{(1)}_{\mu;\alpha}A^{(1)\mu}_{;\beta} \rangle + \kappa^2\langle\mathcal{P}_{\alpha\beta}\rangle 
\end{eqnarray}
where
\begin{eqnarray}
\langle\mathcal{P}_{\alpha\beta}\rangle &=&  - \frac{1}{2}\bar{F}_{\beta}^{\epsilon} \bar{F}_{\epsilon}^{\rho} \langle \psi_{\alpha}\,^{\tau} \psi_{\rho \tau} \rangle - \frac{3}{2}\bar{F}_{\alpha}^{\epsilon} \bar{F}_{\epsilon}^{\rho} \langle \psi_{\beta}\,^{\tau} \psi_{\rho \tau} \rangle \nonumber \\
&&- \frac{1}{8}\bar{F}^{\epsilon \rho} \bar{F}_{\epsilon \rho} \langle \psi_{\beta}\,^{\tau} \psi_{\alpha \tau}\rangle+\frac{1}{2}\bar{F}^{\epsilon \rho} \bar{F}_{\epsilon}^{\tau} \langle \psi_{\alpha \rho} \psi_{\beta \tau} \rangle\nonumber\\
&&+2\bar{F}_{\alpha}^{\epsilon} \bar{F}^{\rho \tau} \langle \psi_{\beta \rho} \psi_{\epsilon \tau} \rangle -\bar{F}^{\epsilon \rho} \bar{F}_{\epsilon}^{\tau} \langle \psi_{\alpha \beta} \psi_{\rho \tau}\rangle \nonumber \\
&&+ \bar{g}_{\alpha\beta}\left[\frac{3}{2}\bar{F}_{\epsilon\rho}\bar{F}_{\mu\tau}\langle \psi^{\epsilon\mu}\psi^{\rho\tau} \rangle-\bar{F}^{\rho}_{\epsilon}\bar{F}^{\tau}_{\rho}\langle \psi^{\epsilon\mu}\psi_{\mu\tau}\rangle\right.\nonumber \\
&&\left.+\frac{1}{8}\left(\bar{F}\right)^2\langle \psi^2 \rangle  \right]
\end{eqnarray}
\section{Details of the effective source term of $ f(R) $ gravity}
\label{app:sourceterm}

For a scalar perturbed energy-momentum tensor given by
\begin{eqnarray}
T^S_{\mu\nu} &\equiv& \left(\begin{array}{c|c}
\tau_{ab}\textbf{S} & r\tau^{(S)}_{a}\textbf{S}_{B}\\
---&---------       \\
r\tau^{(S)}_{a}\textbf{S}_{B} & r^2 \delta P\gamma_{AB}\textbf{S}+ r^2 \tau^{(S)}_{T}\textbf{S}_{AB}
\end{array}\right),
\end{eqnarray}
from \cite{Kodama2004b} the source term for the scalar perturbation of a charged black-hole was found to be
\begin{eqnarray}
S^{eff}_\pm &=& a^S_\pm S_\Phi + b^S_\pm S_\mathcal{A} \\
S_\Phi &=& \frac{g}{rH}\left[ -HS_{T}-\frac{P_1}{H}\frac{S_{t}}{i\omega}-4g\frac{r\left(S_{t}\right)'}{i\omega}-4rgS_{r}\right. \nonumber \\
&& \left.+\frac{P_2}{H}\frac{rS^{r}_{t}}{i\omega}+2r^{2}\frac{\left(S^{r}_{t}\right)'}{i\omega}+2r^{2}S^{r}_{r} \right]  \label{Sphi}\\
S_\mathcal{A} &=& \frac{2\sqrt{2}Qg}{i\omega r^2H}\left(2gS_t-rS^r_t\right) \label{SA}
\end{eqnarray}
where the prime denotes radial derivatives and
\begin{eqnarray}
P_1 &=& -\frac{32Q^4}{r^4}+\frac{48Q^2}{r^2}\left(\frac{2M}{r}-1\right)-\frac{48M^2}{r^2}+\frac{4M}{r}\left(8-k^2\right)\nonumber \\
&&-2k^2\left(k^2-2\right) \\
P_2 &=& -\frac{32Q^2}{r^2}+\frac{24M}{r} \\
S^a_b &=& \kappa^2\tau^a_b \qquad S_a = \frac{r\kappa^2}{k}\tau_a^{(S)} \qquad S_T = \frac{2r^2\kappa^2}{k^2}\tau_T^{(S)}
\end{eqnarray}
Equating $ T^S_{\mu\nu}=T^{eff}_{\mu\nu} $ the components of $ T^S_{\mu\nu} $ were found from \cite{Bhattacharyya2017} in terms of the massive field $ \Phi $, using which components of $ S^a_b $, $ S_a $, and $ S_T $ were found for (\ref{Sphi}) and (\ref{SA}). $ S^{eff}_\pm $ is only relevant around the horizon of the black hole where the presence of $ \Phi $ is at the largest. Hence, the coefficients $ c_\pm $ and $ d_\pm $ of (\ref{effsauce}) were calculated around the horizon in a power series of $ g(y)\equiv g $ around the horizon and only contribution from $ \frac{1}{g} $ is relevant, which turn out to be
\begin{eqnarray}
c_+ &=& -\,{\frac { \left( \frac{3}{2}\,{q}^{2}+{k}^{2}y-2\,y+\frac{1}{2}\,\sqrt {9+9\,{q}^{4}+ \left( -14+16\,{k}^{2} \right) {q}^{2}}+\frac{3}{2} \right) q \left( {q}^{2}y-16\,{q}^{2}+y \right)  \left({q}^{2}y-2\,{q}^{2}+y \right) }{4g{y}^{7}H}}\nonumber\\
\\
d_+ &=& {\frac { \left( \frac{3}{2}\,{q}^{2}+{k}^{2}y-2\,y+\frac{1}{2}\,\sqrt {9+9\,{q}^{4}+ \left( -14+16\,{k}^{2} \right) {q}^{2}}+\frac{3}{2} \right) q \left( {q}^{2}y-2\,{q}^{2}+y \right) }{2{y}^{4}g}} \\
c_-&=& -\frac {3 \left( {q}^{2}y-\frac{8}{3}\,{q}^{2}+\frac{1}{3}\,\sqrt {9+9\,{q}^{4}+\left( -14+16\,{k}^{2} \right) {q}^{2}}y+y \right)  \left( {q}^{2}y-2\,{q}^{2}+y \right)  \left( {q}^{2}y-16\,{q}^{2}+y \right) }{4g{y}^{7}H}\nonumber\\
\\
d_- &=& \frac {3 \left( {q}^{2}y-\frac{8}{3}\,{q}^{2}+\frac{1}{3}\,\sqrt {9+9\,{q}^{4}+\left( -14+16\,{k}^{2} \right) {q}^{2}}y+y \right)  \left( {q}^{2}y-2\,{q}^{2}+y \right) }{2{y}^{4}g}
\end{eqnarray}

\section{Asymptotic behavior of $ h_{\mu\nu}^{V/S} $ and connection of $ \Phi_\pm^{V/S} $ to $ h_{+/\times} $}\label{asympt}
In \cite{Martel2005,Nagar2005,Bhattacharyya2017} it was shown that only the traceless part of $ h_{AB} $ contributes to the radiation escaping to asymptotic infinity, and a connection was found between polarizations $ h_{+/\times} $ and the gauge invariant perturbation variables of a Schwarzschild space-time

In \cite{Nagar2005} it was shown using the tetrad formalism developed in \cite{Newman1962} that at asymptotic infinity, $ h_{AB} $ can be written in the locally flat coordinate system of an observer  as
\begin{eqnarray}\label{tetrad}
h_{\hat{A}\hat{B}} & =& \textbf{e}^A_{\hat{A}}\textbf{e}^B_{\hat{B}}h_{AB}
\end{eqnarray}
where $ \textbf{e}^A_{\hat{A}} = diag\left[r^{-1},(r\sin\theta)^{-1}\right] $ is the observer's local tetrad and $ \hat{A} $ is the tetrad index.
The traceless part of $ h_{AB} $ has the form \cite{Kodama2000}, where an implicit summation over multipole index $ \ell $ and projection index $ m $ was assumed
\begin{eqnarray} \label{trless}
h_{AB} &=& r^2\left(H^S_T\textbf{S}_{AB}+H^V_T\textbf{V}_{AB}\right)
\end{eqnarray}
Using (\ref{tetrad}) in (\ref{trless}) one obtains the traceless part of the perturbation in $ \mathcal{S}^2 $ at a large distance from the black-hole in a locally flat space-time as
\begin{eqnarray}\label{orth}
h_{\hat{A}\hat{B}} &\equiv& H^S_T\left(\begin{array}{c c}
\textbf{S}_{\theta\theta} & \textbf{S}_{\theta\phi} \\ 
\frac{\textbf{S}_{\theta\phi}}{\sin^2\theta} & \frac{\textbf{S}_{\phi\phi}}{\sin^2\theta}
\end{array}\right) + H^V_T\left(\begin{array}{c c}
\textbf{V}_{\theta\theta} & \textbf{V}_{\theta\phi} \\ 
\frac{\textbf{V}_{\theta\phi}}{\sin^2\theta} & \frac{\textbf{V}_{\phi\phi}}{\sin^2\theta}
\end{array}\right) \nonumber\\
\\
&=& \left(\begin{array}{cc} 
h_+ & h_\times \\ 
h_\times & -h_+
\end{array}\right).
\end{eqnarray}
from which polarization amplitudes $ h_{+/\times} $ can be read off from (\ref{orth}) as
\begin{eqnarray}
h_+ & =& \left(H_T^S\textbf{S}_{\theta\theta}+H^V_T\textbf{V}_{\theta\theta}\right) \label{plus}\\
h_\times & =& \frac{1}{\sin^2\theta}\left(H_T^S\textbf{S}_{\theta\phi}+H^V_T\textbf{V}_{\theta\phi}\right). \label{cross}
\end{eqnarray}
The asymptotic relationship between the perturbation variables and the scalar/vector master variables were found from \cite{Martel2005,Nagar2005}, using which (\ref{plus}) and (\ref{cross}) becomes
\begin{eqnarray}
h_+ & =& \frac{1}{r}\left(\Phi_S\textbf{S}_{\theta\theta}+\Phi_V\textbf{V}_{\theta\theta}\right) \label{plusm}\\
h_\times & =& \frac{1}{r\sin^2\theta}\left(\Phi_S\textbf{S}_{\theta\phi}+\Phi_V\textbf{V}_{\theta\phi}\right). \label{crossm}
\end{eqnarray}
which can be inverted as
\begin{eqnarray}
\frac{\Phi_S}{r} &=& \int h_+\textbf{S}_{\theta\theta}d\Omega + \int \sin^2\theta h_\times\textbf{S}_{\theta\phi} d\Omega \label{intp1} \\
\frac{\Phi_V}{r} &=& \int h_+\textbf{V}_{\theta\theta}d\Omega + \int \sin^2\theta h_\times\textbf{V}_{\theta\phi} d\Omega \label{intp2}
\end{eqnarray}
Observation wise, the above relations are impractical since it involves observing the polarization amplitudes at each point of the surface of a sphere and integrating over it - which is unlikely, unless in future the technological challenge of encompassing the entirety of a black hole with detectors can be overcome. Earth bound detectors can only observe gravitational waves on a small patch of the sphere, given which, it is useful to find the quantity in the LHS of (\ref{intp1}) and (\ref{intp2}) per unit solid angle, for which we obtain
\begin{eqnarray}
\frac{d\Psi_S}{d\Omega} &=& h_+\textbf{S}_{\theta\theta} + \sin^2\theta h_\times\textbf{S}_{\theta\phi}\\
\frac{d\Psi_V}{d\Omega} &=& h_+\textbf{V}_{\theta\theta} + \sin^2\theta h_\times\textbf{V}_{\theta\phi}
\end{eqnarray}
where $ \Psi_{V/S} = \frac{\Phi_{V/S}}{r} $.
%

\begin{thebibliography}{58}%
\makeatletter
\providecommand \@ifxundefined [1]{%
 \@ifx{#1\undefined}
}%
\providecommand \@ifnum [1]{%
 \ifnum #1\expandafter \@firstoftwo
 \else \expandafter \@secondoftwo
 \fi
}%
\providecommand \@ifx [1]{%
 \ifx #1\expandafter \@firstoftwo
 \else \expandafter \@secondoftwo
 \fi
}%
\providecommand \natexlab [1]{#1}%
\providecommand \enquote  [1]{``#1''}%
\providecommand \bibnamefont  [1]{#1}%
\providecommand \bibfnamefont [1]{#1}%
\providecommand \citenamefont [1]{#1}%
\providecommand \href@noop [0]{\@secondoftwo}%
\providecommand \href [0]{\begingroup \@sanitize@url \@href}%
\providecommand \@href[1]{\@@startlink{#1}\@@href}%
\providecommand \@@href[1]{\endgroup#1\@@endlink}%
\providecommand \@sanitize@url [0]{\catcode `\\12\catcode `\$12\catcode
  `\&12\catcode `\#12\catcode `\^12\catcode `\_12\catcode `\%12\relax}%
\providecommand \@@startlink[1]{}%
\providecommand \@@endlink[0]{}%
\providecommand \url  [0]{\begingroup\@sanitize@url \@url }%
\providecommand \@url [1]{\endgroup\@href {#1}{\urlprefix }}%
\providecommand \urlprefix  [0]{URL }%
\providecommand \Eprint [0]{\href }%
\providecommand \doibase [0]{http://dx.doi.org/}%
\providecommand \selectlanguage [0]{\@gobble}%
\providecommand \bibinfo  [0]{\@secondoftwo}%
\providecommand \bibfield  [0]{\@secondoftwo}%
\providecommand \translation [1]{[#1]}%
\providecommand \BibitemOpen [0]{}%
\providecommand \bibitemStop [0]{}%
\providecommand \bibitemNoStop [0]{.\EOS\space}%
\providecommand \EOS [0]{\spacefactor3000\relax}%
\providecommand \BibitemShut  [1]{\csname bibitem#1\endcsname}%
\let\auto@bib@innerbib\@empty
\bibitem [{\citenamefont {Abbott}\ \emph {et~al.}(2016)\citenamefont {Abbott}
  \emph {et~al.}}]{Abbott:2016blz}%
  \BibitemOpen
  \bibfield  {author} {\bibinfo {author} {\bibfnamefont {B.~P.}\ \bibnamefont
  {Abbott}} \emph {et~al.} (\bibinfo {collaboration} {Virgo, LIGO
  Scientific}),\ }\href {\doibase 10.1103/PhysRevLett.116.061102} {\bibfield
  {journal} {\bibinfo  {journal} {Phys. Rev. Lett.}\ }\textbf {\bibinfo
  {volume} {116}},\ \bibinfo {pages} {061102} (\bibinfo {year} {2016})},\
  \Eprint {http://arxiv.org/abs/1602.03837} {arXiv:1602.03837 [gr-qc]}
  \BibitemShut {NoStop}%
\bibitem [{\citenamefont {Abbott}\ \emph
  {et~al.}(2017{\natexlab{a}})\citenamefont {Abbott} \emph
  {et~al.}}]{Abbott:2017oio}%
  \BibitemOpen
  \bibfield  {author} {\bibinfo {author} {\bibfnamefont {B.~P.}\ \bibnamefont
  {Abbott}} \emph {et~al.} (\bibinfo {collaboration} {Virgo, LIGO
  Scientific}),\ }\href {\doibase 10.1103/PhysRevLett.119.141101} {\bibfield
  {journal} {\bibinfo  {journal} {Phys. Rev. Lett.}\ }\textbf {\bibinfo
  {volume} {119}},\ \bibinfo {pages} {141101} (\bibinfo {year}
  {2017}{\natexlab{a}})},\ \Eprint {http://arxiv.org/abs/1709.09660}
  {arXiv:1709.09660 [gr-qc]} \BibitemShut {NoStop}%
\bibitem [{\citenamefont {Abbott}\ \emph
  {et~al.}(2017{\natexlab{b}})\citenamefont {Abbott} \emph
  {et~al.}}]{TheLIGOScientific:2017qsa}%
  \BibitemOpen
  \bibfield  {author} {\bibinfo {author} {\bibfnamefont {B.}~\bibnamefont
  {Abbott}} \emph {et~al.} (\bibinfo {collaboration} {Virgo, LIGO
  Scientific}),\ }\href {\doibase 10.1103/PhysRevLett.119.161101} {\bibfield
  {journal} {\bibinfo  {journal} {Phys. Rev. Lett.}\ }\textbf {\bibinfo
  {volume} {119}},\ \bibinfo {pages} {161101} (\bibinfo {year}
  {2017}{\natexlab{b}})},\ \Eprint {http://arxiv.org/abs/1710.05832}
  {arXiv:1710.05832 [gr-qc]} \BibitemShut {NoStop}%
\bibitem [{\citenamefont {Stelle}(1977)}]{Stelle1977}%
  \BibitemOpen
  \bibfield  {author} {\bibinfo {author} {\bibfnamefont {K.~S.}\ \bibnamefont
  {Stelle}},\ }\href {\doibase 10.1103/PhysRevD.16.953} {\bibfield  {journal}
  {\bibinfo  {journal} {Physical Review D}\ }\textbf {\bibinfo {volume} {16}},\
  \bibinfo {pages} {953} (\bibinfo {year} {1977})},\ \Eprint
  {http://arxiv.org/abs/arXiv:1306.6701v1} {arXiv:arXiv:1306.6701v1}
  \BibitemShut {NoStop}%
\bibitem [{\citenamefont {Clifton}\ \emph {et~al.}(2012)\citenamefont
  {Clifton}, \citenamefont {Ferreira}, \citenamefont {Padilla},\ and\
  \citenamefont {Skordis}}]{Clifton2012}%
  \BibitemOpen
  \bibfield  {author} {\bibinfo {author} {\bibfnamefont {T.}~\bibnamefont
  {Clifton}}, \bibinfo {author} {\bibfnamefont {P.~G.}\ \bibnamefont
  {Ferreira}}, \bibinfo {author} {\bibfnamefont {A.}~\bibnamefont {Padilla}}, \
  and\ \bibinfo {author} {\bibfnamefont {C.}~\bibnamefont {Skordis}},\ }\href
  {\doibase 10.1016/j.physrep.2012.01.001} {\bibfield  {journal} {\bibinfo
  {journal} {Physics Reports}\ }\textbf {\bibinfo {volume} {513}},\ \bibinfo
  {pages} {1} (\bibinfo {year} {2012})},\ \Eprint
  {http://arxiv.org/abs/1106.2476} {arXiv:1106.2476} \BibitemShut {NoStop}%
\bibitem [{\citenamefont {Will}(1993)}]{Will:1993ns}%
  \BibitemOpen
  \bibfield  {author} {\bibinfo {author} {\bibfnamefont {C.}~\bibnamefont
  {Will}},\ }\href {https://books.google.co.in/books?id=BhnUITA7sDIC} {\emph
  {\bibinfo {title} {Theory and Experiment in Gravitational Physics}}}\
  (\bibinfo  {publisher} {Cambridge University Press},\ \bibinfo {year}
  {1993})\BibitemShut {NoStop}%
\bibitem [{\citenamefont {De~Felice}\ and\ \citenamefont
  {Tsujikawa}(2010)}]{DeFelice:2010aj}%
  \BibitemOpen
  \bibfield  {author} {\bibinfo {author} {\bibfnamefont {A.}~\bibnamefont
  {De~Felice}}\ and\ \bibinfo {author} {\bibfnamefont {S.}~\bibnamefont
  {Tsujikawa}},\ }\href {\doibase 10.12942/lrr-2010-3} {\bibfield  {journal}
  {\bibinfo  {journal} {Living Rev. Rel.}\ }\textbf {\bibinfo {volume} {13}},\
  \bibinfo {pages} {3} (\bibinfo {year} {2010})},\ \Eprint
  {http://arxiv.org/abs/1002.4928} {arXiv:1002.4928 [gr-qc]} \BibitemShut
  {NoStop}%
\bibitem [{\citenamefont {Sotiriou}\ and\ \citenamefont
  {Faraoni}(2010)}]{RevModPhys.82.451}%
  \BibitemOpen
  \bibfield  {author} {\bibinfo {author} {\bibfnamefont {T.~P.}\ \bibnamefont
  {Sotiriou}}\ and\ \bibinfo {author} {\bibfnamefont {V.}~\bibnamefont
  {Faraoni}},\ }\href {\doibase 10.1103/RevModPhys.82.451} {\bibfield
  {journal} {\bibinfo  {journal} {Rev. Mod. Phys.}\ }\textbf {\bibinfo {volume}
  {82}},\ \bibinfo {pages} {451} (\bibinfo {year} {2010})}\BibitemShut
  {NoStop}%
\bibitem [{\citenamefont {Nojiri}\ and\ \citenamefont
  {Odintsov}(2011)}]{Nojiri:2010wj}%
  \BibitemOpen
  \bibfield  {author} {\bibinfo {author} {\bibfnamefont {S.}~\bibnamefont
  {Nojiri}}\ and\ \bibinfo {author} {\bibfnamefont {S.~D.}\ \bibnamefont
  {Odintsov}},\ }\href {\doibase 10.1016/j.physrep.2011.04.001} {\bibfield
  {journal} {\bibinfo  {journal} {Phys. Rept.}\ }\textbf {\bibinfo {volume}
  {505}},\ \bibinfo {pages} {59} (\bibinfo {year} {2011})},\ \Eprint
  {http://arxiv.org/abs/1011.0544} {arXiv:1011.0544 [gr-qc]} \BibitemShut
  {NoStop}%
\bibitem [{\citenamefont {Nojiri}\ \emph {et~al.}(2017)\citenamefont {Nojiri},
  \citenamefont {Odintsov},\ and\ \citenamefont {Oikonomou}}]{Nojiri:2017ncd}%
  \BibitemOpen
  \bibfield  {author} {\bibinfo {author} {\bibfnamefont {S.}~\bibnamefont
  {Nojiri}}, \bibinfo {author} {\bibfnamefont {S.~D.}\ \bibnamefont
  {Odintsov}}, \ and\ \bibinfo {author} {\bibfnamefont {V.~K.}\ \bibnamefont
  {Oikonomou}},\ }\href {\doibase 10.1016/j.physrep.2017.06.001} {\bibfield
  {journal} {\bibinfo  {journal} {Phys. Rept.}\ }\textbf {\bibinfo {volume}
  {692}},\ \bibinfo {pages} {1} (\bibinfo {year} {2017})},\ \Eprint
  {http://arxiv.org/abs/1705.11098} {arXiv:1705.11098 [gr-qc]} \BibitemShut
  {NoStop}%
\bibitem [{\citenamefont {Regge}\ and\ \citenamefont
  {Wheeler}(1957)}]{Wheeler}%
  \BibitemOpen
  \bibfield  {author} {\bibinfo {author} {\bibfnamefont {T.}~\bibnamefont
  {Regge}}\ and\ \bibinfo {author} {\bibfnamefont {J.~A.}\ \bibnamefont
  {Wheeler}},\ }\href {\doibase 10.1103/PhysRev.108.1063} {\bibfield  {journal}
  {\bibinfo  {journal} {Phys. Rev.}\ }\textbf {\bibinfo {volume} {108}},\
  \bibinfo {pages} {1063} (\bibinfo {year} {1957})}\BibitemShut {NoStop}%
\bibitem [{\citenamefont {Zerilli}(1970{\natexlab{a}})}]{Zerilli1970a}%
  \BibitemOpen
  \bibfield  {author} {\bibinfo {author} {\bibfnamefont {F.~J.}\ \bibnamefont
  {Zerilli}},\ }\href {\doibase 10.1103/PhysRevLett.24.737} {\bibfield
  {journal} {\bibinfo  {journal} {Physical Review Letters}\ }\textbf {\bibinfo
  {volume} {24}},\ \bibinfo {pages} {737} (\bibinfo {year}
  {1970}{\natexlab{a}})}\BibitemShut {NoStop}%
\bibitem [{\citenamefont {Zerilli}(1970{\natexlab{b}})}]{Zerilli1970}%
  \BibitemOpen
  \bibfield  {author} {\bibinfo {author} {\bibfnamefont {F.~J.}\ \bibnamefont
  {Zerilli}},\ }\href {\doibase 10.1103/PhysRevD.2.2141} {\bibfield  {journal}
  {\bibinfo  {journal} {Physical Review D}\ }\textbf {\bibinfo {volume} {2}},\
  \bibinfo {pages} {2141} (\bibinfo {year} {1970}{\natexlab{b}})}\BibitemShut
  {NoStop}%
\bibitem [{\citenamefont {Zerilli}(1974)}]{Zerilli1974}%
  \BibitemOpen
  \bibfield  {author} {\bibinfo {author} {\bibfnamefont {F.~J.}\ \bibnamefont
  {Zerilli}},\ }\href {\doibase 10.1103/PhysRevD.9.860} {\bibfield  {journal}
  {\bibinfo  {journal} {Physical Review D}\ }\textbf {\bibinfo {volume} {9}},\
  \bibinfo {pages} {860} (\bibinfo {year} {1974})}\BibitemShut {NoStop}%
\bibitem [{\citenamefont {Chandrasekhar}(2002)}]{Chandrasekhar:1985kt}%
  \BibitemOpen
  \bibfield  {author} {\bibinfo {author} {\bibfnamefont {S.}~\bibnamefont
  {Chandrasekhar}},\ }\href {https://cds.cern.ch/record/579245} {\emph
  {\bibinfo {title} {{The mathematical theory of black holes}}}},\ Oxford
  classic texts in the physical sciences\ (\bibinfo  {publisher} {Oxford Univ.
  Press},\ \bibinfo {address} {Oxford},\ \bibinfo {year} {2002})\BibitemShut
  {NoStop}%
\bibitem [{\citenamefont {Vishveshwara}(1970)}]{Vishveshwara1970a}%
  \BibitemOpen
  \bibfield  {author} {\bibinfo {author} {\bibfnamefont {C.~V.}\ \bibnamefont
  {Vishveshwara}},\ }\href {\doibase 10.1103/PhysRevD.1.2870} {\bibfield
  {journal} {\bibinfo  {journal} {Physical Review D}\ }\textbf {\bibinfo
  {volume} {1}},\ \bibinfo {pages} {2870} (\bibinfo {year} {1970})}\BibitemShut
  {NoStop}%
\bibitem [{\citenamefont {Mino}\ \emph {et~al.}(1997)\citenamefont {Mino},
  \citenamefont {Sasaki}, \citenamefont {Shibata}, \citenamefont {Tagoshi},\
  and\ \citenamefont {Tanaka}}]{Mino:1997bx}%
  \BibitemOpen
  \bibfield  {author} {\bibinfo {author} {\bibfnamefont {Y.}~\bibnamefont
  {Mino}}, \bibinfo {author} {\bibfnamefont {M.}~\bibnamefont {Sasaki}},
  \bibinfo {author} {\bibfnamefont {M.}~\bibnamefont {Shibata}}, \bibinfo
  {author} {\bibfnamefont {H.}~\bibnamefont {Tagoshi}}, \ and\ \bibinfo
  {author} {\bibfnamefont {T.}~\bibnamefont {Tanaka}},\ }\href {\doibase
  10.1143/PTPS.128.1} {\bibfield  {journal} {\bibinfo  {journal} {Prog. Theor.
  Phys. Suppl.}\ }\textbf {\bibinfo {volume} {128}},\ \bibinfo {pages} {1}
  (\bibinfo {year} {1997})},\ \Eprint {http://arxiv.org/abs/gr-qc/9712057}
  {arXiv:gr-qc/9712057 [gr-qc]} \BibitemShut {NoStop}%
\bibitem [{\citenamefont {Sasaki}\ and\ \citenamefont
  {Tagoshi}(2003)}]{Sasaki:2003xr}%
  \BibitemOpen
  \bibfield  {author} {\bibinfo {author} {\bibfnamefont {M.}~\bibnamefont
  {Sasaki}}\ and\ \bibinfo {author} {\bibfnamefont {H.}~\bibnamefont
  {Tagoshi}},\ }\href {\doibase 10.12942/lrr-2003-6} {\bibfield  {journal}
  {\bibinfo  {journal} {Living Rev. Rel.}\ }\textbf {\bibinfo {volume} {6}},\
  \bibinfo {pages} {6} (\bibinfo {year} {2003})},\ \Eprint
  {http://arxiv.org/abs/gr-qc/0306120} {arXiv:gr-qc/0306120 [gr-qc]}
  \BibitemShut {NoStop}%
\bibitem [{\citenamefont {Kokkotas}\ and\ \citenamefont
  {Schmidt}(1999)}]{Kokkotas:1999bd}%
  \BibitemOpen
  \bibfield  {author} {\bibinfo {author} {\bibfnamefont {K.~D.}\ \bibnamefont
  {Kokkotas}}\ and\ \bibinfo {author} {\bibfnamefont {B.~G.}\ \bibnamefont
  {Schmidt}},\ }\href {\doibase 10.12942/lrr-1999-2} {\bibfield  {journal}
  {\bibinfo  {journal} {Living Rev. Rel.}\ }\textbf {\bibinfo {volume} {2}},\
  \bibinfo {pages} {2} (\bibinfo {year} {1999})},\ \Eprint
  {http://arxiv.org/abs/gr-qc/9909058} {arXiv:gr-qc/9909058 [gr-qc]}
  \BibitemShut {NoStop}%
\bibitem [{\citenamefont {Berti}\ \emph {et~al.}(2009)\citenamefont {Berti},
  \citenamefont {Cardoso},\ and\ \citenamefont {Starinets}}]{Berti:2009kk}%
  \BibitemOpen
  \bibfield  {author} {\bibinfo {author} {\bibfnamefont {E.}~\bibnamefont
  {Berti}}, \bibinfo {author} {\bibfnamefont {V.}~\bibnamefont {Cardoso}}, \
  and\ \bibinfo {author} {\bibfnamefont {A.~O.}\ \bibnamefont {Starinets}},\
  }\href {\doibase 10.1088/0264-9381/26/16/163001} {\bibfield  {journal}
  {\bibinfo  {journal} {Class. Quant. Grav.}\ }\textbf {\bibinfo {volume}
  {26}},\ \bibinfo {pages} {163001} (\bibinfo {year} {2009})},\ \Eprint
  {http://arxiv.org/abs/0905.2975} {arXiv:0905.2975 [gr-qc]} \BibitemShut
  {NoStop}%
\bibitem [{\citenamefont {Nollert}(1999)}]{0264-9381-16-12-201}%
  \BibitemOpen
  \bibfield  {author} {\bibinfo {author} {\bibfnamefont {H.-P.}\ \bibnamefont
  {Nollert}},\ }\href {http://stacks.iop.org/0264-9381/16/i=12/a=201}
  {\bibfield  {journal} {\bibinfo  {journal} {Classical and Quantum Gravity}\
  }\textbf {\bibinfo {volume} {16}},\ \bibinfo {pages} {R159} (\bibinfo {year}
  {1999})}\BibitemShut {NoStop}%
\bibitem [{\citenamefont {Konoplya}\ and\ \citenamefont
  {Zhidenko}(2011)}]{Konoplya:2011qq}%
  \BibitemOpen
  \bibfield  {author} {\bibinfo {author} {\bibfnamefont {R.~A.}\ \bibnamefont
  {Konoplya}}\ and\ \bibinfo {author} {\bibfnamefont {A.}~\bibnamefont
  {Zhidenko}},\ }\href {\doibase 10.1103/RevModPhys.83.793} {\bibfield
  {journal} {\bibinfo  {journal} {Rev. Mod. Phys.}\ }\textbf {\bibinfo {volume}
  {83}},\ \bibinfo {pages} {793} (\bibinfo {year} {2011})},\ \Eprint
  {http://arxiv.org/abs/1102.4014} {arXiv:1102.4014 [gr-qc]} \BibitemShut
  {NoStop}%
\bibitem [{\citenamefont {Berry}\ and\ \citenamefont {Gair}(2011)}]{Berry2011}%
  \BibitemOpen
  \bibfield  {author} {\bibinfo {author} {\bibfnamefont {C.~P.~L.}\
  \bibnamefont {Berry}}\ and\ \bibinfo {author} {\bibfnamefont {J.~R.}\
  \bibnamefont {Gair}},\ }\href {\doibase 10.1103/PhysRevD.83.104022}
  {\bibfield  {journal} {\bibinfo  {journal} {Physical Review D}\ }\textbf
  {\bibinfo {volume} {83}} (\bibinfo {year} {2011}),\
  10.1103/PhysRevD.83.104022},\ \Eprint {http://arxiv.org/abs/1104.0819}
  {arXiv:1104.0819} \BibitemShut {NoStop}%
\bibitem [{\citenamefont {Capozziello}\ \emph {et~al.}(2008)\citenamefont
  {Capozziello}, \citenamefont {Corda},\ and\ \citenamefont {{De
  Laurentis}}}]{Capozziello2008}%
  \BibitemOpen
  \bibfield  {author} {\bibinfo {author} {\bibfnamefont {S.}~\bibnamefont
  {Capozziello}}, \bibinfo {author} {\bibfnamefont {C.}~\bibnamefont {Corda}},
  \ and\ \bibinfo {author} {\bibfnamefont {M.~F.}\ \bibnamefont {{De
  Laurentis}}},\ }\href {\doibase 10.1016/j.physletb.2008.10.001} {\bibfield
  {journal} {\bibinfo  {journal} {Physics Letters, Section B: Nuclear,
  Elementary Particle and High-Energy Physics}\ }\textbf {\bibinfo {volume}
  {669}},\ \bibinfo {pages} {255} (\bibinfo {year} {2008})},\ \Eprint
  {http://arxiv.org/abs/0812.2272} {arXiv:0812.2272} \BibitemShut {NoStop}%
\bibitem [{\citenamefont {Will}(1994)}]{PhysRevD.50.6058}%
  \BibitemOpen
  \bibfield  {author} {\bibinfo {author} {\bibfnamefont {C.~M.}\ \bibnamefont
  {Will}},\ }\href {\doibase 10.1103/PhysRevD.50.6058} {\bibfield  {journal}
  {\bibinfo  {journal} {Phys. Rev. D}\ }\textbf {\bibinfo {volume} {50}},\
  \bibinfo {pages} {6058} (\bibinfo {year} {1994})}\BibitemShut {NoStop}%
\bibitem [{\citenamefont {Myung}(2016)}]{Myung2016}%
  \BibitemOpen
  \bibfield  {author} {\bibinfo {author} {\bibfnamefont {Y.~S.}\ \bibnamefont
  {Myung}},\ }\href {\doibase 10.1155/2016/3901734} {\bibfield  {journal}
  {\bibinfo  {journal} {Adv. High Energy Phys.}\ }\textbf {\bibinfo {volume}
  {2016}},\ \bibinfo {pages} {3901734} (\bibinfo {year} {2016})},\ \Eprint
  {http://arxiv.org/abs/1608.01764} {arXiv:1608.01764} \BibitemShut {NoStop}%
\bibitem [{\citenamefont {Capozziello}\ \emph
  {et~al.}(2017{\natexlab{a}})\citenamefont {Capozziello}, \citenamefont
  {De~Laurentis}, \citenamefont {Nojiri},\ and\ \citenamefont
  {Odintsov}}]{Capozziello:2017vdi}%
  \BibitemOpen
  \bibfield  {author} {\bibinfo {author} {\bibfnamefont {S.}~\bibnamefont
  {Capozziello}}, \bibinfo {author} {\bibfnamefont {M.}~\bibnamefont
  {De~Laurentis}}, \bibinfo {author} {\bibfnamefont {S.}~\bibnamefont
  {Nojiri}}, \ and\ \bibinfo {author} {\bibfnamefont {S.~D.}\ \bibnamefont
  {Odintsov}},\ }\href {\doibase 10.1103/PhysRevD.95.083524} {\bibfield
  {journal} {\bibinfo  {journal} {Phys. Rev.}\ }\textbf {\bibinfo {volume}
  {D95}},\ \bibinfo {pages} {083524} (\bibinfo {year} {2017}{\natexlab{a}})},\
  \Eprint {http://arxiv.org/abs/1702.05517} {arXiv:1702.05517 [gr-qc]}
  \BibitemShut {NoStop}%
\bibitem [{\citenamefont {Nojiri}\ and\ \citenamefont
  {Odintsov}(2014)}]{Nojiri:2014jqa}%
  \BibitemOpen
  \bibfield  {author} {\bibinfo {author} {\bibfnamefont {S.}~\bibnamefont
  {Nojiri}}\ and\ \bibinfo {author} {\bibfnamefont {S.~D.}\ \bibnamefont
  {Odintsov}},\ }\href {\doibase 10.1016/j.physletb.2014.06.070} {\bibfield
  {journal} {\bibinfo  {journal} {Phys. Lett.}\ }\textbf {\bibinfo {volume}
  {B735}},\ \bibinfo {pages} {376} (\bibinfo {year} {2014})},\ \Eprint
  {http://arxiv.org/abs/1405.2439} {arXiv:1405.2439 [gr-qc]} \BibitemShut
  {NoStop}%
\bibitem [{\citenamefont {Nojiri}\ and\ \citenamefont
  {Odintsov}(2017)}]{Nojiri:2017kex}%
  \BibitemOpen
  \bibfield  {author} {\bibinfo {author} {\bibfnamefont {S.}~\bibnamefont
  {Nojiri}}\ and\ \bibinfo {author} {\bibfnamefont {S.~D.}\ \bibnamefont
  {Odintsov}},\ }\href {\doibase 10.1103/PhysRevD.96.104008} {\bibfield
  {journal} {\bibinfo  {journal} {Phys. Rev.}\ }\textbf {\bibinfo {volume}
  {D96}},\ \bibinfo {pages} {104008} (\bibinfo {year} {2017})},\ \Eprint
  {http://arxiv.org/abs/1708.05226} {arXiv:1708.05226 [hep-th]} \BibitemShut
  {NoStop}%
\bibitem [{\citenamefont {{De La Cruz-Dombriz}}\ \emph
  {et~al.}(2009)\citenamefont {{De La Cruz-Dombriz}}, \citenamefont {Dobado},\
  and\ \citenamefont {Maroto}}]{DeLaCruz-Dombriz2009}%
  \BibitemOpen
  \bibfield  {author} {\bibinfo {author} {\bibfnamefont {A.}~\bibnamefont {{De
  La Cruz-Dombriz}}}, \bibinfo {author} {\bibfnamefont {A.}~\bibnamefont
  {Dobado}}, \ and\ \bibinfo {author} {\bibfnamefont {A.~L.}\ \bibnamefont
  {Maroto}},\ }\href {\doibase 10.1103/PhysRevD.80.124011} {\bibfield
  {journal} {\bibinfo  {journal} {Physical Review D - Particles, Fields,
  Gravitation and Cosmology}\ }\textbf {\bibinfo {volume} {80}} (\bibinfo
  {year} {2009}),\ 10.1103/PhysRevD.80.124011},\ \Eprint
  {http://arxiv.org/abs/0907.3872} {arXiv:0907.3872} \BibitemShut {NoStop}%
\bibitem [{\citenamefont {Bhattacharyya}\ and\ \citenamefont
  {Shankaranarayanan}(2017)}]{Bhattacharyya2017}%
  \BibitemOpen
  \bibfield  {author} {\bibinfo {author} {\bibfnamefont {S.}~\bibnamefont
  {Bhattacharyya}}\ and\ \bibinfo {author} {\bibfnamefont {S.}~\bibnamefont
  {Shankaranarayanan}},\ }\href {\doibase 10.1103/PhysRevD.96.064044}
  {\bibfield  {journal} {\bibinfo  {journal} {Phys. Rev. D}\ }\textbf {\bibinfo
  {volume} {96}} (\bibinfo {year} {2017}),\ 10.1103/PhysRevD.96.064044},\
  \Eprint {http://arxiv.org/abs/1704.07044} {arXiv:1704.07044} \BibitemShut
  {NoStop}%
\bibitem [{\citenamefont {Woodard}(2007)}]{Woodard2007}%
  \BibitemOpen
  \bibfield  {author} {\bibinfo {author} {\bibfnamefont {R.}~\bibnamefont
  {Woodard}},\ }\href {\doibase 10.1007/978-3-540-71013-4_14} {\bibfield
  {journal} {\bibinfo  {journal} {The Invisible Universe: Dark Matter and Dark
  Energy}\ ,\ \bibinfo {pages} {1}} (\bibinfo {year} {2007})},\ \Eprint
  {http://arxiv.org/abs/0601672v2} {arXiv:0601672v2 [arXiv:astro-ph]}
  \BibitemShut {NoStop}%
\bibitem [{\citenamefont {Starobinsky}(1980)}]{Starobinsky1980}%
  \BibitemOpen
  \bibfield  {author} {\bibinfo {author} {\bibfnamefont {A.~A.}\ \bibnamefont
  {Starobinsky}},\ }\href {\doibase 10.1016/0370-2693(80)90670-X} {\bibfield
  {journal} {\bibinfo  {journal} {Physics Letters B}\ }\textbf {\bibinfo
  {volume} {91}},\ \bibinfo {pages} {99} (\bibinfo {year} {1980})}\BibitemShut
  {NoStop}%
\bibitem [{\citenamefont {Zwiebach}(1985)}]{Zwiebach:1985uq}%
  \BibitemOpen
  \bibfield  {author} {\bibinfo {author} {\bibfnamefont {B.}~\bibnamefont
  {Zwiebach}},\ }\href {\doibase 10.1016/0370-2693(85)91616-8} {\bibfield
  {journal} {\bibinfo  {journal} {Phys. Lett.}\ }\textbf {\bibinfo {volume}
  {156B}},\ \bibinfo {pages} {315} (\bibinfo {year} {1985})}\BibitemShut
  {NoStop}%
\bibitem [{\citenamefont {Tseytlin}(1986)}]{TSEYTLIN198692}%
  \BibitemOpen
  \bibfield  {author} {\bibinfo {author} {\bibfnamefont {A.}~\bibnamefont
  {Tseytlin}},\ }\href {\doibase https://doi.org/10.1016/0370-2693(86)90930-5}
  {\bibfield  {journal} {\bibinfo  {journal} {Physics Letters B}\ }\textbf
  {\bibinfo {volume} {176}},\ \bibinfo {pages} {92 } (\bibinfo {year}
  {1986})}\BibitemShut {NoStop}%
\bibitem [{\citenamefont {Oh}(1986)}]{Oh:1985bm}%
  \BibitemOpen
  \bibfield  {author} {\bibinfo {author} {\bibfnamefont {P.}~\bibnamefont
  {Oh}},\ }\href {\doibase 10.1016/0370-2693(86)90802-6} {\bibfield  {journal}
  {\bibinfo  {journal} {Phys. Lett.}\ }\textbf {\bibinfo {volume} {166B}},\
  \bibinfo {pages} {292} (\bibinfo {year} {1986})}\BibitemShut {NoStop}%
\bibitem [{\citenamefont {Charmousis}(2009)}]{Charmousis:2008kc}%
  \BibitemOpen
  \bibfield  {author} {\bibinfo {author} {\bibfnamefont {C.}~\bibnamefont
  {Charmousis}},\ }\bibfield  {booktitle} {\emph {\bibinfo {booktitle}
  {{Proceedings, 4th Aegean Summer School: Black Holes: Mytilene, Island of
  Lesvos, Greece, September 17-22, 2007}}},\ }\href {\doibase
  10.1007/978-3-540-88460-6_8} {\bibfield  {journal} {\bibinfo  {journal}
  {Lect. Notes Phys.}\ }\textbf {\bibinfo {volume} {769}},\ \bibinfo {pages}
  {299} (\bibinfo {year} {2009})},\ \Eprint {http://arxiv.org/abs/0805.0568}
  {arXiv:0805.0568 [gr-qc]} \BibitemShut {NoStop}%
\bibitem [{\citenamefont {Kodama}\ \emph {et~al.}(2000)\citenamefont {Kodama},
  \citenamefont {Ishibashi},\ and\ \citenamefont {Seto}}]{Kodama2000}%
  \BibitemOpen
  \bibfield  {author} {\bibinfo {author} {\bibfnamefont {H.}~\bibnamefont
  {Kodama}}, \bibinfo {author} {\bibfnamefont {A.}~\bibnamefont {Ishibashi}}, \
  and\ \bibinfo {author} {\bibfnamefont {O.}~\bibnamefont {Seto}},\ }\href@noop
  {} {\  (\bibinfo {year} {2000})},\ \Eprint {http://arxiv.org/abs/0004160v3}
  {arXiv:0004160v3 [arXiv:hep-th]} \BibitemShut {NoStop}%
\bibitem [{\citenamefont {Kodama}\ and\ \citenamefont
  {Ishibashi}(2003)}]{Kodama2003}%
  \BibitemOpen
  \bibfield  {author} {\bibinfo {author} {\bibfnamefont {H.}~\bibnamefont
  {Kodama}}\ and\ \bibinfo {author} {\bibfnamefont {A.}~\bibnamefont
  {Ishibashi}},\ }\href {\doibase 10.1143/PTP.110.701} {\bibfield  {journal}
  {\bibinfo  {journal} {Progress of Theoretical Physics}\ }\textbf {\bibinfo
  {volume} {110}},\ \bibinfo {pages} {701} (\bibinfo {year} {2003})},\ \Eprint
  {http://arxiv.org/abs/0305147} {arXiv:0305147 [hep-th]} \BibitemShut
  {NoStop}%
\bibitem [{\citenamefont {Kodama}(2004)}]{Kodama2004}%
  \BibitemOpen
  \bibfield  {author} {\bibinfo {author} {\bibfnamefont {H.}~\bibnamefont
  {Kodama}},\ }\href@noop {} {\  (\bibinfo {year} {2004})},\ \Eprint
  {http://arxiv.org/abs/0403030v2} {arXiv:0403030v2 [arXiv:hep-th]}
  \BibitemShut {NoStop}%
\bibitem [{\citenamefont {Kumar}\ and\ \citenamefont
  {Shankaranarayanan}(2016)}]{Kumar:2015bha}%
  \BibitemOpen
  \bibfield  {author} {\bibinfo {author} {\bibfnamefont {S.~S.}\ \bibnamefont
  {Kumar}}\ and\ \bibinfo {author} {\bibfnamefont {S.}~\bibnamefont
  {Shankaranarayanan}},\ }\href {\doibase 10.1140/epjc/s10052-016-4241-3}
  {\bibfield  {journal} {\bibinfo  {journal} {Eur. Phys. J.}\ }\textbf
  {\bibinfo {volume} {C76}},\ \bibinfo {pages} {400} (\bibinfo {year}
  {2016})},\ \Eprint {http://arxiv.org/abs/1504.00501} {arXiv:1504.00501
  [quant-ph]} \BibitemShut {NoStop}%
\bibitem [{\citenamefont {Kodama}\ and\ \citenamefont
  {Ishibashi}(2004)}]{Kodama2004b}%
  \BibitemOpen
  \bibfield  {author} {\bibinfo {author} {\bibfnamefont {H.}~\bibnamefont
  {Kodama}}\ and\ \bibinfo {author} {\bibfnamefont {A.}~\bibnamefont
  {Ishibashi}},\ }\href {https://arxiv.org/pdf/hep-th/0308128.pdf} {\
  (\bibinfo {year} {2004})},\ \Eprint {http://arxiv.org/abs/0308128v4}
  {arXiv:0308128v4 [arXiv:hep-th]} \BibitemShut {NoStop}%
\bibitem [{\citenamefont {Gunter}(1980)}]{Gunter1980}%
  \BibitemOpen
  \bibfield  {author} {\bibinfo {author} {\bibfnamefont {D.~L.}\ \bibnamefont
  {Gunter}},\ }\href {www.jstor.org
  http://rsta.royalsocietypublishing.org/content/296/1422/497.short} {\bibfield
   {journal} {\bibinfo  {journal} {Philosophical Transactions of the Royal
  Society of London. Series A}\ }\textbf {\bibinfo {volume} {296}},\ \bibinfo
  {pages} {497} (\bibinfo {year} {1980})}\BibitemShut {NoStop}%
\bibitem [{\citenamefont {Isaacson}(1968{\natexlab{a}})}]{Isaacson1968b}%
  \BibitemOpen
  \bibfield  {author} {\bibinfo {author} {\bibfnamefont {R.~A.}\ \bibnamefont
  {Isaacson}},\ }\href {\doibase 10.1103/PhysRev.166.1263} {\bibfield
  {journal} {\bibinfo  {journal} {Physical Review}\ }\textbf {\bibinfo {volume}
  {166}},\ \bibinfo {pages} {1263} (\bibinfo {year}
  {1968}{\natexlab{a}})}\BibitemShut {NoStop}%
\bibitem [{\citenamefont {Isaacson}(1968{\natexlab{b}})}]{Isaacson1968a}%
  \BibitemOpen
  \bibfield  {author} {\bibinfo {author} {\bibfnamefont {R.~A.}\ \bibnamefont
  {Isaacson}},\ }\href {\doibase 10.1103/PhysRev.166.1272} {\bibfield
  {journal} {\bibinfo  {journal} {Physical Review}\ }\textbf {\bibinfo {volume}
  {166}},\ \bibinfo {pages} {1272} (\bibinfo {year}
  {1968}{\natexlab{b}})}\BibitemShut {NoStop}%
\bibitem [{\citenamefont {Aguirregabiria}\ \emph {et~al.}(1996)\citenamefont
  {Aguirregabiria}, \citenamefont {Chamorro},\ and\ \citenamefont
  {Virbhadra}}]{Aguirregabiria:1995qz}%
  \BibitemOpen
  \bibfield  {author} {\bibinfo {author} {\bibfnamefont {J.~M.}\ \bibnamefont
  {Aguirregabiria}}, \bibinfo {author} {\bibfnamefont {A.}~\bibnamefont
  {Chamorro}}, \ and\ \bibinfo {author} {\bibfnamefont {K.~S.}\ \bibnamefont
  {Virbhadra}},\ }\href {\doibase 10.1007/BF02109529} {\bibfield  {journal}
  {\bibinfo  {journal} {Gen. Rel. Grav.}\ }\textbf {\bibinfo {volume} {28}},\
  \bibinfo {pages} {1393} (\bibinfo {year} {1996})},\ \Eprint
  {http://arxiv.org/abs/gr-qc/9501002} {arXiv:gr-qc/9501002 [gr-qc]}
  \BibitemShut {NoStop}%
\bibitem [{\citenamefont {Virbhadra}(1990)}]{Virbhadra:1990vs}%
  \BibitemOpen
  \bibfield  {author} {\bibinfo {author} {\bibfnamefont {K.~S.}\ \bibnamefont
  {Virbhadra}},\ }\href {\doibase 10.1103/PhysRevD.41.1086} {\bibfield
  {journal} {\bibinfo  {journal} {Phys. Rev.}\ }\textbf {\bibinfo {volume}
  {D41}},\ \bibinfo {pages} {1086} (\bibinfo {year} {1990})}\BibitemShut
  {NoStop}%
\bibitem [{\citenamefont {Capozziello}\ \emph
  {et~al.}(2017{\natexlab{b}})\citenamefont {Capozziello}, \citenamefont
  {Capriolo},\ and\ \citenamefont {Transirico}}]{Capozziello:2017xla}%
  \BibitemOpen
  \bibfield  {author} {\bibinfo {author} {\bibfnamefont {S.}~\bibnamefont
  {Capozziello}}, \bibinfo {author} {\bibfnamefont {M.}~\bibnamefont
  {Capriolo}}, \ and\ \bibinfo {author} {\bibfnamefont {M.}~\bibnamefont
  {Transirico}},\ }\href {\doibase 10.1002/andp.201600376} {\bibfield
  {journal} {\bibinfo  {journal} {Annalen Phys.}\ }\textbf {\bibinfo {volume}
  {529}},\ \bibinfo {pages} {1600376} (\bibinfo {year} {2017}{\natexlab{b}})},\
  \Eprint {http://arxiv.org/abs/1702.01162} {arXiv:1702.01162 [gr-qc]}
  \BibitemShut {NoStop}%
\bibitem [{\citenamefont {Nzioki}\ \emph {et~al.}(2014)\citenamefont {Nzioki},
  \citenamefont {Goswami},\ and\ \citenamefont {Dunsby}}]{Nzioki2014}%
  \BibitemOpen
  \bibfield  {author} {\bibinfo {author} {\bibfnamefont {A.~M.}\ \bibnamefont
  {Nzioki}}, \bibinfo {author} {\bibfnamefont {R.}~\bibnamefont {Goswami}}, \
  and\ \bibinfo {author} {\bibfnamefont {P.~K.~S.}\ \bibnamefont {Dunsby}},\
  }\href {http://arxiv.org/abs/1408.0152} {\ ,\ \bibinfo {pages} {29} (\bibinfo
  {year} {2014})},\ \Eprint {http://arxiv.org/abs/1408.0152} {arXiv:1408.0152}
  \BibitemShut {NoStop}%
\bibitem [{\citenamefont {Casseres}(2017)}]{10.1088/1361-6382/aa8e2e}%
  \BibitemOpen
  \bibfield  {author} {\bibinfo {author} {\bibfnamefont {P.~M.~C.}\
  \bibnamefont {Casseres}},\ }\href
  {http://iopscience.iop.org/10.1088/1361-6382/aa8e2e} {\bibfield  {journal}
  {\bibinfo  {journal} {Classical and Quantum Gravity}\ } (\bibinfo {year}
  {2017})}\BibitemShut {NoStop}%
\bibitem [{\citenamefont {Skakala}\ and\ \citenamefont
  {Shankaranarayanan}(2014)}]{Skakala:2014gca}%
  \BibitemOpen
  \bibfield  {author} {\bibinfo {author} {\bibfnamefont {J.}~\bibnamefont
  {Skakala}}\ and\ \bibinfo {author} {\bibfnamefont {S.}~\bibnamefont
  {Shankaranarayanan}},\ }\href {\doibase 10.1088/0264-9381/31/17/175005}
  {\bibfield  {journal} {\bibinfo  {journal} {Class. Quant. Grav.}\ }\textbf
  {\bibinfo {volume} {31}},\ \bibinfo {pages} {175005} (\bibinfo {year}
  {2014})},\ \Eprint {http://arxiv.org/abs/1402.6166} {arXiv:1402.6166 [gr-qc]}
  \BibitemShut {NoStop}%
\bibitem [{\citenamefont {Pani}\ \emph
  {et~al.}(2012{\natexlab{a}})\citenamefont {Pani}, \citenamefont {Cardoso},
  \citenamefont {Gualtieri}, \citenamefont {Berti},\ and\ \citenamefont
  {Ishibashi}}]{Pani:2012bp}%
  \BibitemOpen
  \bibfield  {author} {\bibinfo {author} {\bibfnamefont {P.}~\bibnamefont
  {Pani}}, \bibinfo {author} {\bibfnamefont {V.}~\bibnamefont {Cardoso}},
  \bibinfo {author} {\bibfnamefont {L.}~\bibnamefont {Gualtieri}}, \bibinfo
  {author} {\bibfnamefont {E.}~\bibnamefont {Berti}}, \ and\ \bibinfo {author}
  {\bibfnamefont {A.}~\bibnamefont {Ishibashi}},\ }\href {\doibase
  10.1103/PhysRevD.86.104017} {\bibfield  {journal} {\bibinfo  {journal} {Phys.
  Rev.}\ }\textbf {\bibinfo {volume} {D86}},\ \bibinfo {pages} {104017}
  (\bibinfo {year} {2012}{\natexlab{a}})},\ \Eprint
  {http://arxiv.org/abs/1209.0773} {arXiv:1209.0773 [gr-qc]} \BibitemShut
  {NoStop}%
\bibitem [{\citenamefont {Pani}\ \emph
  {et~al.}(2012{\natexlab{b}})\citenamefont {Pani}, \citenamefont {Cardoso},
  \citenamefont {Gualtieri}, \citenamefont {Berti},\ and\ \citenamefont
  {Ishibashi}}]{Pani:2012vp}%
  \BibitemOpen
  \bibfield  {author} {\bibinfo {author} {\bibfnamefont {P.}~\bibnamefont
  {Pani}}, \bibinfo {author} {\bibfnamefont {V.}~\bibnamefont {Cardoso}},
  \bibinfo {author} {\bibfnamefont {L.}~\bibnamefont {Gualtieri}}, \bibinfo
  {author} {\bibfnamefont {E.}~\bibnamefont {Berti}}, \ and\ \bibinfo {author}
  {\bibfnamefont {A.}~\bibnamefont {Ishibashi}},\ }\href {\doibase
  10.1103/PhysRevLett.109.131102} {\bibfield  {journal} {\bibinfo  {journal}
  {Phys. Rev. Lett.}\ }\textbf {\bibinfo {volume} {109}},\ \bibinfo {pages}
  {131102} (\bibinfo {year} {2012}{\natexlab{b}})},\ \Eprint
  {http://arxiv.org/abs/1209.0465} {arXiv:1209.0465 [gr-qc]} \BibitemShut
  {NoStop}%
\bibitem [{\citenamefont {Ferrari}\ and\ \citenamefont
  {Mashhoon}(1984)}]{Ferrari1984}%
  \BibitemOpen
  \bibfield  {author} {\bibinfo {author} {\bibfnamefont {V.}~\bibnamefont
  {Ferrari}}\ and\ \bibinfo {author} {\bibfnamefont {B.}~\bibnamefont
  {Mashhoon}},\ }\href {\doibase 10.1103/PhysRevD.30.295} {\bibfield  {journal}
  {\bibinfo  {journal} {Physical Review D}\ }\textbf {\bibinfo {volume} {30}},\
  \bibinfo {pages} {295} (\bibinfo {year} {1984})}\BibitemShut {NoStop}%
\bibitem [{\citenamefont {Misner}\ \emph {et~al.}(1973)\citenamefont {Misner},
  \citenamefont {Thorne},\ and\ \citenamefont {Wheeler}}]{Misner:1974qy}%
  \BibitemOpen
  \bibfield  {author} {\bibinfo {author} {\bibfnamefont {C.~W.}\ \bibnamefont
  {Misner}}, \bibinfo {author} {\bibfnamefont {K.~S.}\ \bibnamefont {Thorne}},
  \ and\ \bibinfo {author} {\bibfnamefont {J.~A.}\ \bibnamefont {Wheeler}},\
  }\href@noop {} {\emph {\bibinfo {title} {{Gravitation}}}}\ (\bibinfo
  {publisher} {W. H. Freeman},\ \bibinfo {address} {San Francisco},\ \bibinfo
  {year} {1973})\BibitemShut {NoStop}%
\bibitem [{\citenamefont {Martel}\ and\ \citenamefont
  {Poisson}(2005)}]{Martel2005}%
  \BibitemOpen
  \bibfield  {author} {\bibinfo {author} {\bibfnamefont {K.}~\bibnamefont
  {Martel}}\ and\ \bibinfo {author} {\bibfnamefont {E.}~\bibnamefont
  {Poisson}},\ }\href {\doibase 10.1103/PhysRevD.71.104003} {\bibfield
  {journal} {\bibinfo  {journal} {Physical Review D}\ }\textbf {\bibinfo
  {volume} {71}},\ \bibinfo {pages} {104003} (\bibinfo {year} {2005})},\
  \Eprint {http://arxiv.org/abs/0502028} {arXiv:0502028 [gr-qc]} \BibitemShut
  {NoStop}%
\bibitem [{\citenamefont {Nagar}\ and\ \citenamefont
  {Rezzolla}(2005)}]{Nagar2005}%
  \BibitemOpen
  \bibfield  {author} {\bibinfo {author} {\bibfnamefont {A.}~\bibnamefont
  {Nagar}}\ and\ \bibinfo {author} {\bibfnamefont {L.}~\bibnamefont
  {Rezzolla}},\ }\href {\doibase 10.1088/0264-9381/22/16/R01} {\bibfield
  {journal} {\bibinfo  {journal} {Class. Quantum Grav}\ }\textbf {\bibinfo
  {volume} {22}},\ \bibinfo {pages} {167} (\bibinfo {year} {2005})},\ \Eprint
  {http://arxiv.org/abs/0502064} {arXiv:0502064 [gr-qc]} \BibitemShut {NoStop}%
\bibitem [{\citenamefont {Newman}\ and\ \citenamefont
  {Penrose}(1962)}]{Newman1962}%
  \BibitemOpen
  \bibfield  {author} {\bibinfo {author} {\bibfnamefont {E.~T.}\ \bibnamefont
  {Newman}}\ and\ \bibinfo {author} {\bibfnamefont {R.}~\bibnamefont
  {Penrose}},\ }\href@noop {} {\bibfield  {journal} {\bibinfo  {journal}
  {Journal of Mathematical Physics}\ }\textbf {\bibinfo {volume} {3}},\
  \bibinfo {pages} {566} (\bibinfo {year} {1962})}\BibitemShut {NoStop}%
\end{thebibliography}
%

\end{document}